\documentclass{aa}  

\usepackage{amssymb}
\usepackage{pifont}
\usepackage{graphicx}
\usepackage{xcolor}
\usepackage{txfonts}

\begin{document}

   \title{The interstellar medium of high-redshift galaxies: Gathering clues from \ion{C}{III}] and [\ion{C}{II}] lines}

   \titlerunning{ISM of high-$z$ galaxies.}
   \author{V. Markov
          \inst{1}, S. Carniani\inst{1}, L. Vallini\inst{1}, A. Ferrara\inst{1},  A. Pallottini\inst{1}, R. Maiolino\inst{2,}\inst{3,}\inst{4}, S. Gallerani\inst{1}, L. Pentericci\inst{5}}
   \authorrunning{Markov et al. (2022)}
   \institute{Scuola Normale Superiore, Piazza dei Cavalieri 7, 56126 Pisa, Italy\\
              \email{vladan.markov@sns.it}
         \and
        Kavli Institute for Cosmology, University of Cambridge, Madingley Road, Cambridge, CB3 OHA, UK
        \and 
        Cavendish Laboratory - Astrophysics Group, University of Cambridge, 19 JJ Thompson Avenue, Cambridge, CB3 OHE, UK
        \and 
        Department of Physics and Astronomy, University College London, Gower Street, London WC1E 6BT, UK
        \and
        INAF – Osservatorio Astronomico di Roma, via Frascati 33, 00078, Monteporzio Catone, Italy
             }

   \date{Received February 16, 2022; accepted June 05, 2022}

  \abstract
   {A tight relation between [\ion{C}{II}] $ 158 \ {\rm{\mu m}}$ line luminosity and the star formation rate (SFR) has been observed for local galaxies. At high redshift ($z > 5$), galaxies instead deviate downwards from the local $\Sigma_{\rm{[\ion{C}{II}]}}-\Sigma_{\rm{SFR}}$ relation. This deviation  might be caused by different interstellar medium (ISM) properties in galaxies at early epochs.}
   {To test this hypothesis, we combined the [\ion{C}{II}] and SFR data with \ion{C}{III}] $1909\AA$ line observations and our physical models. We additionally investigated how ISM properties, such as burstiness, $\kappa_s$, total gas density, $n$, and metallicity, $Z$, affect the deviation from the $\Sigma_{\rm{[\ion{C}{II}]}}-\Sigma_{\rm{SFR}}$ relation in these sources. 
   }
   {We present the VLT/X-SHOOTER observations targeting the \ion{C}{III}] $\lambda1909$ line emission in three galaxies at $5.5<z<7.0$. We include archival X-SHOOTER data of two other sources at $5.5<z<7.0$ and the VLT/MUSE archival data of six galaxies at $z \sim 2$. We extend our sample of galaxies with eleven star-forming systems at $6 < z< 7.5,$ with either \ion{C}{III}] or [\ion{C}{II}] detection reported in the literature.}
   {We detected \ion{C}{III}] $\lambda \lambda1907, 1909$ line emission in HZ10 and we derived the intrinsic, integrated flux of the $\ion{C}{III}]\ \lambda1909$ line. We constrained the ISM properties for our sample of galaxies, $\kappa_s$, $n$, and $Z,$ by applying our physically motivated model based on the MCMC algorithm. For the most part, high-$z$ star-forming galaxies show subsolar metallicities. The majority of the sources have $\log{(\kappa_s)} \gtrsim 1$, that is, they overshoot the Kennicutt-Schmidt (KS) relation by about one order of magnitude.}
   { Our findings suggest that the whole KS relation might be shifted upwards at early times. Furthermore, all the high-$z$ galaxies of our sample lie below the $\Sigma_{\rm{[\ion{C}{II}]}}-\Sigma_{\rm{SFR}}$ local relation. The total gas density, $n,$ shows the strongest correlation with the deviation from the local $\Sigma_{\rm{[\ion{C}{II}]}}-\Sigma_{\rm{SFR}}$ relation, namely, low-density high-$z$ systems have lower [\ion{C}{II}] surface brightness, in agreement with theoretical models.}


   \maketitle
%

\section{Introduction}

The Epoch of Reionization (EoR) is an important period in cosmic history when the formation of stars in the first galaxies played a key role in the complete reionization of the Universe by $z \sim 6$ and in the metal enrichment of the intergalactic medium (IGM; see \citealt{2006ARA&A..44..415F}; \citealt{2015PASA...32...45B}; \citealt{2016ARA&A..54..761S}; \citealt{finkelstein_2016}; and \citealt{2018PhR...780....1D} for reviews). 
Characterizing the internal properties of early galaxies, namely, the physical properties of their interstellar medium (ISM) is fundamental to our understanding of galaxy formation and evolution. This involves building theoretical models (see, e.g., \citealt{2018PhR...780....1D} for a review) and constraining these models with observations tracing the different phases of the ISM (see \citealt{2016ARA&A..54..761S} and \citealt{finkelstein_2016} for a review).

In recent years, ALMA observations of the [\ion{C}{II}] $^2P_{3/2} - ^2P_{1/2}$ fine-structure line emission at a rest-frame wavelength of $\lambda  = 157.74 \ {\rm{\mu m}}$ have been widely used to characterize the ISM physical conditions: gas density, metallicity, and relative abundance of different gas phases of $z>6$ star-forming galaxies. It can be also used to put constraints on the global properties such as the galaxy morphology, redshift, and size (\citealt{2005A&A...440L..51M, 2009A&A...500L...1M}; \citealt{2013A&A...559A..29C, 2017A&A...605A..42C,  2018MNRAS.478.1170C, 2018ApJ...854L...7C}; \citealt{2015Natur.522..455C}; \citealt{2015ApJ...807..180W}; \citealt{2016ApJ...829L..11P}; \citealt{2017ApJ...836L...2B}; \citealt{2017ApJ...851..145M, 2019ApJ...881..124M}; \citealt{2018Natur.553..178S}; see \citealt{2013ARA&A..51..105C} for a review). The [\ion{C}{II}] line is among the brightest emission lines in the rest-frame far-infrared (FIR) band (\citealt{1991ApJ...373..423S, 2010ApJ...724..957S}) and mainly traces the so-called photo-dissociation regions (PDRs; \citealt{2022arXiv220205867W}); this refers, namely, to the dense, mostly neutral external layers of molecular clouds connected to \ion{H}{II} regions. The [\ion{C}{II}] can also originate, albeit to a lesser extent, from the diffuse cold and warm neutral medium  as well as from the mildly diffuse ionized gas (\citealt{2003ApJ...587..278W}; \citealt{2015ApJ...813...36V}; \citealt{2017MNRAS.471.4128P}), which is confirmed by observations  (\citealt{1997ApJ...483..200M};  \citealt{2010ApJ...724..957S}; \citealt{2013A&A...554A.103P}; \citealt{2015A&A...578A..53C}).

The [\ion{C}{II}] emissivity is correlated to the far-ultraviolet (FUV) emission from young OB stars (\citealt{1999RvMP...71..173H}). For this reason, it can be used as tracer of the star formation activity (\citealt{2015ApJ...813...36V}) as confirmed by local observations (\citealt{2011MNRAS.416.2712D, 2014A&A...568A..62D}; \citealt{2015ApJ...800....1H}). However, while at low-$z,$ the luminosity of [\ion{C}{II}] is tightly correlated to star formation rate (SFR), the L$_{[\ion{C}{II}]}$-SFR relation in the distant Universe ($z>4$) has an intrinsic dispersion that is up to $\approx 2\times$ larger than the local one (\citealt{2018MNRAS.478.1170C}; \citealt{2018A&A...609A.130L}; \citealt{2020A&A...643A...3S}; \citealt{2020ApJ...905..102L}),  with several galaxies showing  [\ion{C}{II}] luminosities lower than expected from the local L$_{[\ion{C}{II}]}$-SFR relation (\citealt{2015ApJ...813...36V}; \citealt{2017MNRAS.471.4128P}; \citealt{2017ApJ...846..105O}; \citealt{2018MNRAS.478.1170C}; \citealt{2018A&A...609A.130L}; \citealt{2019MNRAS.482.4906P}; \citealt{2020ApJ...905..102L}). This larger scatter is indicative of a broader range of properties spanned by early galaxies with respect to the local population. Lower [\ion{C}{II}] luminosity in high-$z$ sources can be explained by a low gas metallicity (\citealt{2015ApJ...813...36V}; \citealt{2017ApJ...846..105O}; \citealt{2018MNRAS.478.1170C}; \citealt{2018A&A...609A.130L}; \citealt{2020ApJ...905..102L}), a low molecular gas mass fraction (\citealt{2017ApJ...846..105O}), a low molecular gas content (\citealt{2020ApJ...905..102L}), a strong external pressure acting on molecular clouds (\citealt{2019MNRAS.482.4906P}), and a high intensity of the radiation field owing to the starburst nature of galaxies at the EoR, likely placing these sources well above the  Kennicutt-Schmidt $\Sigma_{\rm SFR}-\Sigma_{\rm gas}$  relation (hereafter, KS; \citealt{1959ApJ...129..243S}; \citealt{1998ApJ...498..541K}), \citep{2015ApJ...813...36V, 2018A&A...609A.130L, 2018MNRAS.478.1170C, 2019MNRAS.487.1689P}.

While most of the above-mentioned studies have concentrated on the integrated $\rm{[\ion{C}{II}]-SFR}$ relation, that is, the $\rm{L_{[\ion{C}{II}]}-SFR}$ relation, in this work, we focus on the spatially resolved $\Sigma_{[\ion{C}{II}]}$-$\Sigma_{\rm{SFR}}$ relation, since recent studies have shown that the  [\ion{C}{II}] surface brightness and SFR per unit area are more closely related to the local properties of the ISM than the integrated values (e.g., \citealt{2014A&A...568A..62D}; \citealt{2020MNRAS.494.5542R}). Almost all high-$z$ galaxies are located below the $\Sigma_{[\ion{C}{II}]}$-$\Sigma_{\rm{SFR}}$ relation in local sources (\citealt{2018MNRAS.478.1170C}). The origin of the lower [\ion{C}{II}] surface brightness in high-$z$ galaxies lies in the overall different local ISM conditions at high-$z$ compared to the local Universe. For instance, lower gas metallicity $Z$ leads to lower carbon abundance and, consequently, to lower $\Sigma_{[\ion{C}{II}]}$ (\citealt{2018MNRAS.478.1170C}; \citealt{2019MNRAS.489....1F}). Next, intense radiation fields in high-$z$ starburst galaxies tend to ionize most of the neutral phase of the ISM, from which the bulk of the [\ion{C}{II}] emission originates \citep{2019MNRAS.487.1689P, 2019MNRAS.489....1F}. Finally,  systems with particularly low density of the [\ion{C}{II}] emitting gas, $n < 10^2 \ \rm{cm^{-3}}$, will have a low $\Sigma_{[\ion{C}{II}]}$ as a reusult of their extended [\ion{C}{II}] emitting regions caused, for instance, by outflows \citep{2018MNRAS.478.1170C} or due to cosmic microwave background (CMB) suppression effects (\citealt{2019MNRAS.487.3007K}).

\cite{2019MNRAS.489....1F} developed a physically motivated, analytical model to explain the $\Sigma_{[\ion{C}{II}]}$-$\Sigma_{\rm{SFR}}$ relation both at low and high redshifts. 
These authors found that the deviation from the local $\Sigma_{\rm [\ion{C}{II}]} - \Sigma_{\rm SFR}$ relation depends on three ISM parameters: the total density of the emitting gas $n$, the gas metallicity $Z$, and the so-called "burstiness"\ parameter, $\kappa_s$, quantifying the efficiency of a galaxy to convert its gas into stars. This physically motivated model and interpretation is supported by results obtained with cosmological simulations (\citealt{2019MNRAS.487.1689P}).

The [\ion{C}{II}] line alone is not sufficient to resolve the degeneracy between the three ISM properties.  
Therefore, to break the degeneracy, we need to combine observations of the [\ion{C}{II}] line with an additional emission line tracing the ISM (\citealt{2019MNRAS.489....1F}). Emission lines often used in the literature are the rest-frame FIR fine structure ($^3P_{1} - ^3P_{0}$) transition of the [\ion{O}{III}] emission line at $\lambda = 88 \ \rm{\mu m}$ 
(\citealt{2016Sci...352.1559I}; \citealt{2017A&A...605A..42C}; \citealt{2017MNRAS.467.1300V, 2021MNRAS.505.5543V}; \citealt{2019MNRAS.487.1689P, 2022MNRAS.513.5621P};  \citealt{2020ApJ...896...93H}) and the rotational transitions of the CO molecule (e.g., \citealt{2018MNRAS.473..271V}, \citealt{2019ApJ...882..168P}, \citealt{2021A&A...652A..66P}). However, constraining the ISM properties exploiting the [\ion{O}{III}] is limited to a very specific and limited redshift ranges because of the poor atmospheric transmission at short wavelengths in the ALMA bands. On the other hand, characterizing the $\kappa_s$ parameter requires measurements of the molecular gas content, which are difficult to obtain at high-$z$ because CO observations call for long exposures (e.g., \citealt{2016ApJ...833...70D}; \citealt{2019ApJ...882..168P}).
Another caveat is that molecular gas estimation from the CO emission requires the CO-to-$\rm{H_2}$ conversion factor $\alpha_{\rm{CO}}$, which is highly uncertain -- and even more so at higher redshift (\citealt{2012MNRAS.421.3127N}; \citealt{2020A&A...641A..22M}; see \citealt{2013ARA&A..51..207B}; \citealt{2013ARA&A..51..105C}; and \citealt{2018A&ARv..26....5C} for a review).

Recently, \cite{2020MNRAS.495L..22V}  showed that the rest-frame UV, semi-forbidden line \ion{C}{III}] $\lambda1909$, combined with the rest-frame FIR [\ion{C}{II}] can be a possible alternative to constrain the ISM properties. Apart from the Ly$\alpha$ line, the \ion{C}{III}] $\lambda \lambda1907, 1909$ doublet is typically one of the strongest nebular emission lines in $z > 2$ galaxies  (\citealt{2003ApJ...588...65S}; \citealt{2014MNRAS.445.3200S}; \citealt{2022A&A...659A..16L}). It falls in the near-infrared (NIR) wavelength range at $z > 5$ and it is detectable with ground-based facilities such as the Very Large Telescope (VLT; \citealt{2015MNRAS.450.1846S}; \citealt{2015ApJ...808..139S, 2019MNRAS.482.2422S}) and the Keck Observatory (\citealt{2015ApJ...808..139S}; \citealt{2015MNRAS.450.1846S, 2017MNRAS.464..469S}; \citealt{2017ApJ...851...40L}; \citealt{Hutchison_2019}). Although \ion{C}{III}] detection from high-$z$ galaxies remain challenging, in the near future, this will change considerably thanks to the {\it James Webb Space Telescope} ({\it JWST}) (\citealt{2019MNRAS.483.2621C}). \\

The goal of this paper is to use \ion{C}{III}] and [\ion{C}{II}] emission data to constrain the gas metallicity, $Z$, the burstiness parameter, $\kappa_s$, and the gas density, $n,$ of high-$z$ galaxies by exploiting the \cite{2020MNRAS.495L..22V} model. Furthermore, we want to investigate the connection between the derived ISM properties and the deviation from the local $\Sigma_{[\ion{C}{II}]}$-$\Sigma_{\rm{SFR}}$ relation observed in the early Universe. 
To achieve this goal, we present the new VLT/X-SHOOTER observations targeting \ion{C}{III}] emission in three $z >5$ galaxies. Next, we include the X-SHOOTER archival data of two other $z >5$ galaxies and the VLT/MUSE archival data of six intermediate-$z$ galaxies at $z \sim 2$. We combine \ion{C}{III}] line observations with the [\ion{C}{II}] line emission and SFR measurements, either derived in this work or published in the literature. Finally, to have a representative sample of galaxies for our analysis, we also include galaxies at $6<z<7.5$ with well studied \ion{C}{III}] and [\ion{C}{II}] emission from the literature. 

This paper is organized as follows: a brief overview of the physical models we use in this work to constrain the ISM properties of high-$z$ galaxies is given in Section \ref{model}, the galaxy sample and new observations are presented in Section \ref{Data}, with the reduction of the new and archival data outlined in Section \ref{reduction}. The analysis of the \ion{C}{III}] emission from the novel and archival X-SHOOTER data is presented in Section \ref{xshooter_analysis}. In Section \ref{Results}, we gather our results, while their discussion is presented in Section \ref{Discussion}. The summary and conclusions of this work are outlined in Section \ref{Conclusions}. Throughout this paper we assume the following cosmological parameters: $\Omega_{\rm M} = 0.308$, $\Omega_{\rm \Lambda} = 0.685$, and $h = 0.678$ (\citealt{2016A&A...594A..13P}). We adopt a Kroupa initial mass function (IMF) (\citealt{2001MNRAS.322..231K}), \cite{2012ARA&A..50..531K} scaling relations and, when necessary, we convert the results from the literature with different scaling relations and IMF values.

\section{Physical model} \label{model}

The physically motivated, analytical model we use in this paper was  developed by \cite{2019MNRAS.489....1F}, based on the radiative transfer of both ionizing and non-ionizing UV photons between the different phases of the ISM. The ISM is modeled as a plane-parallel slab with an ionized layer (\ion{H}{II} region), where carbon emission is predominantly in \ion{C}{III}], a neutral (atomic) layer (PDR region), where carbon emits in [\ion{C}{II}], and finally, a molecular gas region where carbon is in a neutral state [\ion{C}{I}]. The [\ion{C}{II}] line surface brightness emerging from the \ion{H}{II} region and PDR can be written as (Eqs. 35 and 32 of \citealt{2019MNRAS.489....1F})\footnote{For a complete
derivation of the equations see \cite{2019MNRAS.489....1F}.}:

\begin{equation}
\Sigma_{\rm [\ion{C}{II}]} \propto n D N_d \ln{(1+ 10^5 w U)},
\end{equation}%
where $n$ is the total gas density in units of $cm^{-3}$, $D$ is the dust-to-gas ratio in units of the Milky Way value (assumed to be proportional to the gas metallicity $Z$ in solar units), $N_d = 1.7 \times 10^{21} D^{-1} \ cm^{-2}$ is the dust column density, $w = 1/(1+0.9D^{0.5})$ is the probability for a Lyman-Werner photon to be absorbed by the dust, and $U = 1.7 \times 10^{14} \Sigma_{\rm SFR}/\Sigma_{\rm gas}^{2}$ is the ionization parameter, which can also be expressed as per Eq. 40 of \citealt{2019MNRAS.489....1F}:

\begin{equation}
U \simeq 10^{-3} k_s^{10/7} \Sigma_{\rm SFR}^{-3/7},
\end{equation}%
where $\kappa_s$ is the "burstiness"\ parameter, first introduced by \cite{2019MNRAS.489....1F}. The $\kappa_s$ parameter is a parametrization of the deviation from the KS relation (\citealt{1998ApJ...498..541K}), which can be expressed as:

\begin{equation}
\Sigma_{\rm SFR} = 10^{-12} \kappa_s \Sigma_{\rm gas}^{1.4}.
\end{equation}%
 The KS relation is an observed power-law correlation between the SFR per unit area $\Sigma_{\rm{SFR}}$ and  molecular gas surface density $\Sigma_{\rm{gas}}$. The $k_s$ parameter can be considered as a proxy of the star formation efficiency. Overall, for "normal" star-forming galaxies that lie on the KS relation the $k_s$ parameter is $k_s = 1$, whereas starburst galaxies located well above the KS relation display $k_s \gg 1$.

According to the \cite{2019MNRAS.489....1F} analytical model, the $\Sigma_{\rm [\ion{C}{II}]}$ of a galaxy with a given $\Sigma_{\rm SFR}$ can be expressed as a function of three ISM parameters: $k_s$, $n$, and $Z$. In general, galaxies with high $\Sigma_{\rm{SFR}}$ tend to experience higher $\Sigma_{[\ion{C}{II}]}$. However, at $z>5$, the $\Sigma_{[\ion{C}{II}]}-\Sigma_{\rm{SFR}}$ relation is more complex than a simple power law and depends on the ISM properties of galaxies. Moreover, for sources with high SFR per unit area $\Sigma_{\rm{SFR}} \gtrsim 1 \ M_\odot \rm{ \ yr^{-1} \ kpc^{-2}}$, [\ion{C}{II}] surface brightness stops depending on SFR and metallicity $Z$, and begins to saturate at $\rm{\log(\Sigma_{[\ion{C}{II}]}/(L_{\odot} \ kpc^{-2}))} \sim 7$. In this regime, $\Sigma_{[\ion{C}{II}]}$ still depends on the gas density $n$ and weakly on ionization parameter $U$ i.e., the burstiness parameter $k_s$. Moreover, for starburst galaxies with $\kappa_s \gg 1$, the model predicts that $\Sigma_{[\ion{C}{II}]}$ is shifted towards higher $\Sigma_{\rm{SFR}}$, placing these objects below the local $\Sigma_{[\ion{C}{II}]}-\Sigma_{\rm{SFR}}$ relation (Fig. 5d in \citealt{2019MNRAS.489....1F}).

\cite{2020MNRAS.495L..22V}  further expanded upon the \citet{2019MNRAS.489....1F} model and developed a statistical method based on a Bayesian approach and a Markov chain Monte Carlo (MCMC) algorithm using the three observed properties: $\Sigma_{\ion{C}{III}]}$, $\Sigma_{[\ion{C}{II}]}$, and the deviation from the local $\Sigma_{[\ion{C}{II}]}-\Sigma_{\rm{SFR}}$ relation $\Delta$. These three observed quantities  depend on the three unknown ISM properties, the burstiness $\kappa_s$, the density $n$, and the metallicity $Z$. For instance, the predicted surface brightness of the [\ion{C}{II}] line $\Sigma_{[\ion{C}{II}]}$ of a source with a fixed $\Sigma_{\rm{SFR}}$ can be expressed in the following way (Eq. (9) of  \citealt{2020MNRAS.495L..22V}):

\begin{equation} \label{sigma_cii}
\Sigma_{\rm [\ion{C}{II}]} = 2.4 \times 10^9 F_{\rm [\ion{C}{II}]} (\Sigma_{\rm SFR}|n, Z, k_s).
\end{equation}%
The surface brightness of the \ion{C}{III}] line $\Sigma_{\ion{C}{III}]}$ can be written in a corresponding way as: $\Sigma_{\rm \ion{C}{III}]} = 2.4 \times 10^9 F_{\rm \ion{C}{III}]} (\Sigma_{\rm SFR}|n, Z, k_s)$ as per Eq. 12 of \citealt{2020MNRAS.495L..22V}. Finally, the deviation from the local $\Sigma_{[\ion{C}{II}]}-\Sigma_{\rm{SFR}}$ relation of a high-$z$ galaxy at fixed $\Sigma_{\rm{SFR}}$, can be defined as (Eq. 14 of \citealt{2020MNRAS.495L..22V})

\begin{equation} \label{delta}
    \rm{\Delta(\Sigma_{\rm SFR}|n, Z, k_s)} = \rm{\log{(\Sigma_{[\ion{C}{II}]}^{\rm{obs}})}} - \rm{\log{(\Sigma_{[\ion{C}{II}]}^{\rm{DL}})}}, 
\end{equation}%
where $\Sigma_{[\ion{C}{II}]}^{\rm{obs}}$  is the observed [\ion{C}{II}] surface brightness and $\Sigma_{[\ion{C}{II}]}^{\rm{DL}}$ is the [\ion{C}{II}] surface brightness predicted from the local $\Sigma_{[\ion{C}{II}]}-\Sigma_{\rm{SFR}}$ relation by \cite{2014A&A...568A..62D}. \citet{2020MNRAS.495L..22V} use the MCMC to fit the three observables: $\Sigma_{\ion{C}{III}]}$, $\Sigma_{[\ion{C}{II}]}$, and $\Delta$, and obtain the posterior probability distribution for the unknown parameters: $\kappa_s$, $n$, and $Z$ -- thereby obtaining the best-fit parameters and the $1\sigma$ uncertainties. 

\section{Sample and spectroscopic observations} \label{Data}

\begin{figure}
\centering
\includegraphics[width=\hsize]{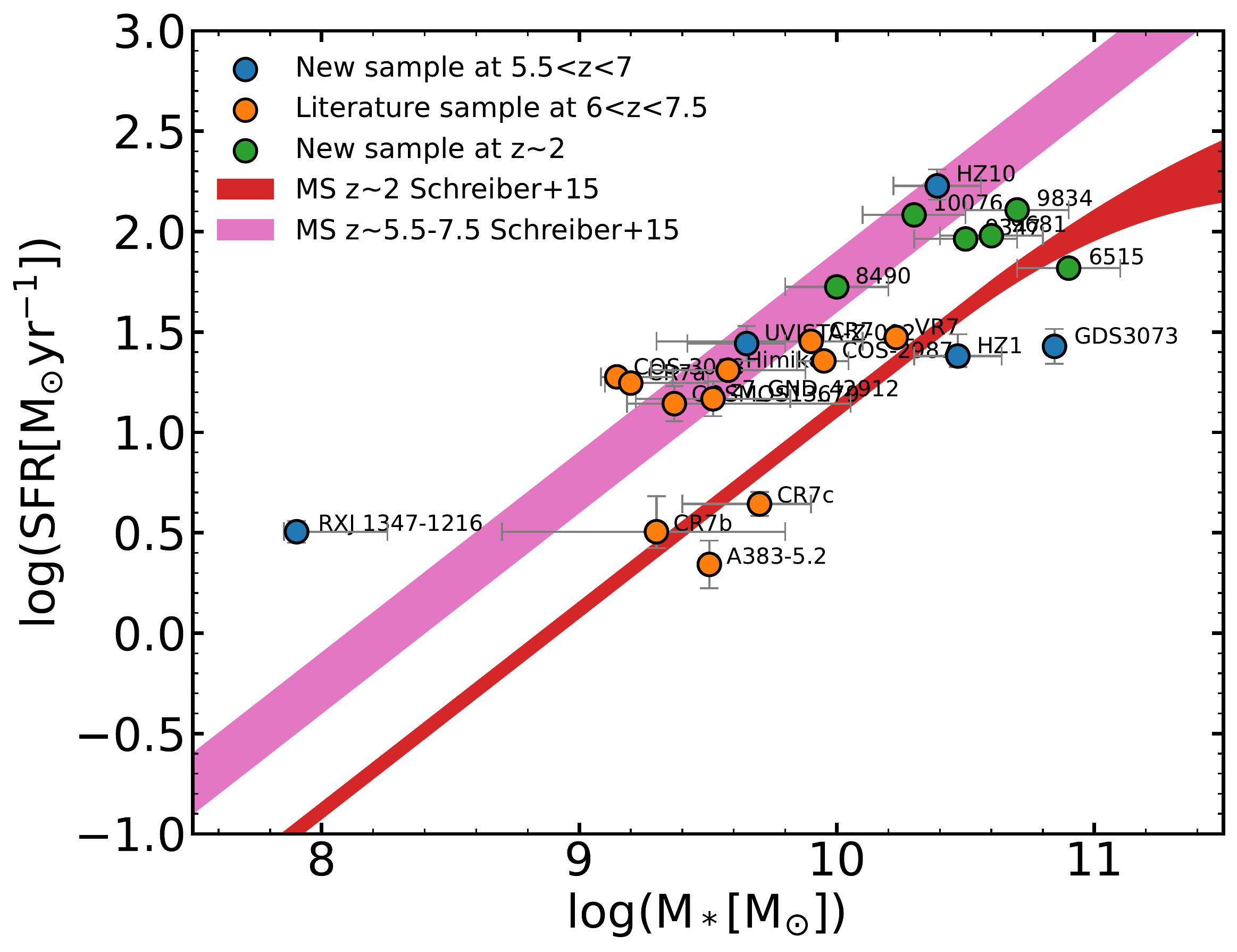}
 \caption{SFR vs. stellar mass of our full galaxy sample. Sources at $5.5 < z < 7.5$ with new and archival \ion{C}{III]} data, galaxies at $z \sim 2$ with archival \ion{C}{III]} data, and literature sample at $6.0 < z < 7.5$ are shown as blue, green, and orange circles, respectively. The best-fit relations for the main sequence (MS) at $5.5 < z < 7.5$ and $1.7<z<2.0$ from \cite{2015A&A...575A..74S} are shown as magenta and red bands, respectively. SFR and stellar mass references: HZ1, HZ10 - \cite{2015Natur.522..455C}; RXJ 1347-1216 - \cite{2016ApJ...817...11H}, \cite{2017ApJ...836L...2B}; GDS3073 - \cite{2020ApJ...897...94G}; UVISTA-Z-002 - \cite{2017MNRAS.466.3612B}, \cite{2021arXiv210613719B}; COS-3018 - \cite{2018Natur.553..178S}; A383-5.2 - \cite{2019ApJ...881..124M}, \cite{2015MNRAS.450.1846S}; z7\_GND\_42912 - \cite{2015A&A...574A..19S}; Himiko - \cite{2018ApJ...854L...7C}, \cite{2020MNRAS.494.1071G}; CR7, CR7a, CR7b, CR7c - \cite{2017ApJ...851..145M}, \cite{2014MNRAS.440.2810B, 2017MNRAS.469..448B}; COS-13679 - \cite{2019ApJ...881..124M}, \cite{2017ApJ...851...40L}; COS-2987 - \cite{2018Natur.553..178S}, \cite{2017ApJ...851...40L}; VR7 - \cite{2019ApJ...881..124M}, \cite{2017MNRAS.472..772M}; 9347, 6515, 10076, 9834, 9681, 8490 - this work, \cite{2018MNRAS.481.1976Z}. 
          }
     \label{Full_sample2}
\end{figure}

For the purpose of this work, we selected star-forming galaxies at $z>2$ with rest-frame UV observations of the \ion{C}{III}] line and rest-frame FIR observations of the [\ion{C}{II}] line. We drew our sample mainly from the literature and archival data: VLT/X-SHOOTER, VLT/MUSE, and ALMA, which have not been published yet. We retrieve the measurements and data of the carbon lines only for those galaxies with a detection in at least one of the two carbon lines. In addition to the public data, we also include three [\ion{C}{II}]-emitting galaxies at z>5 recently observed with X-SHOOTER by our group (PI. S. Carniani). A list of all sources used in this work is reported in Table \ref{detection}. In Fig. \ref{Full_sample2}, we present SFR and stellar mass of our entire sample of galaxies from the literature.

\begin{table}[h]
 \caption[]{\label{detection} Overview of the full galaxy sample used in this paper, grouped by \ion{C}{III}] and [\ion{C}{II}] detection.}
\begin{tabular}{lcccccccc}
 \hline \hline
  \noalign{\smallskip}
  ID & $z_{\rm{sp}}$ &\ion{C}{III}] & [\ion{C}{II}] & Ref. \\
   \noalign{\smallskip}
 \hline
 \noalign{\smallskip}
HZ10 & 5.657 & {\color{blue}  \ding{51}} & {\color{blue} \ding{51}} & [$\dag$, 1]\\
 \noalign{\smallskip}
COS-3018 & 6.854 &{\color{blue} \ding{51}} & {\color{blue} \ding{51}} & [2, 11] \\
 \noalign{\smallskip}
A383-5.2 & 6.029 & {\color{blue} \ding{51}} & {\color{blue} \ding{51}} & [3, 12]\\ %
 \noalign{\smallskip}
 z7\_GND\_42912 & 7.506 & {\color{blue} \ding{51}} & {\color{red} \ding{55}} & [7, 13]\\ %
  \noalign{\smallskip}
HZ1 & 5.689 & {\color{red} \ding{55}} & {\color{blue} \ding{51}} & [$\dag$, 1]\\
 \noalign{\smallskip}
RXJ 1347-1216 & 6.766 &{\color{red} \ding{55}} & {\color{blue} \ding{51}} & [$\dag$, 14] \\
 \noalign{\smallskip}
GDS3073 & 5.563 & {\color{red} \ding{55}} & {\color{blue} \ding{51}} & [$\dag$, 15]\\ %
 \noalign{\smallskip}
 UVISTA-Z-002 & 6.634$^a$ &{\color{red} \ding{55}} & {\color{blue} \ding{51}} & [$\dag$, $\dag$]\\ %
 \noalign{\smallskip}
 Himiko & 6.595 & {\color{red} \ding{55}} & {\color{blue} \ding{51}} & [4, 8]\\
 \noalign{\smallskip}
CR7 & 6.604 &{\color{red} \ding{55}} & {\color{blue} \ding{51}} & [9, 10] \\
 \noalign{\smallskip}
CR7a & 6.604 & {\color{red} \ding{55}} & {\color{blue} \ding{51}} & [9, 10]\\ %
 \noalign{\smallskip}
 CR7b &6.604 & {\color{red} \ding{55}} & {\color{blue} \ding{51}} & [9, 10]\\ %
 \noalign{\smallskip}
 CR7c & 6.604 &{\color{red} \ding{55}} & {\color{blue} \ding{51}} & [9, 10]\\
 \noalign{\smallskip}
COS-13679 & 7.145 &{\color{red} \ding{55}} & {\color{blue} \ding{51}} & [2, 16] \\
 \noalign{\smallskip}
COS-2987  & 6.816 & {\color{red} \ding{55}} & {\color{blue} \ding{51}} & [2, 11]\\ %
 \noalign{\smallskip}
 VR7 & 6.532 &{\color{red} \ding{55}} & {\color{blue} \ding{51}} & [5, 17]\\ %
 \noalign{\smallskip}
 9347 & 1.851$^b$ &{\color{red} \ding{55}} & {\color{blue} \ding{51}} & [$\dag$, 6]\\
 \noalign{\smallskip}
6515 & 1.844$^b$ &{\color{red} \ding{55}} & {\color{blue} \ding{51}} & [$\dag$, 6] \\
 \noalign{\smallskip}
10076 &1.946$^b$ & {\color{red} \ding{55}} & {\color{blue} \ding{51}} & [$\dag$, 6]\\ %
 \noalign{\smallskip}
 9834 & 1.764$^b$ & {\color{red} \ding{55}} & {\color{blue} \ding{51}} & [$\dag$, 6]\\ %
 \noalign{\smallskip}
 9681  & 1.885$^c$ &{\color{red} \ding{55}} & {\color{blue} \ding{51}} & [$\dag$, 6]\\ %
 \noalign{\smallskip}
 8490 & 1.906$^c$ & {\color{red} \ding{55}} & {\color{blue} \ding{51}} & [$\dag$, 6]\\ %
 \noalign{\smallskip}
\hline
\end{tabular}
\tablefoot{Spectroscopic redshift is based on the Ly$\alpha$ emission or spectroscopic Lyman break except when otherwise noted. $^a$ Spectroscopic redshift measured from a Gaussian fit to the integrated spectrum of the [\ion{C}{II}] line from this work. $^b$ Spectroscopic redshift from the [\ion{C}{II}] line emission. $^c$ Spectroscopic redshift from the optical spectra. Typical redshift uncertainties are $\Delta z_{Ly\alpha} \sim 0.0005-0.004$, $\Delta z_{[\ion{C}{II}]} \sim 0.0002-0.0006$, and $\Delta z_{\rm{opt}} \sim 0.001$. The mark {\color{blue} \ding{51}} indicates that the emission line has been detected. The mark {\color{red} \ding{55}} indicates that the emission line has not been detected and an upper limit has been placed on the line flux.}
\tablebib{\ion{C}{III}] and [\ion{C}{II}] line references:   [$\dag$] This work, [1] \cite{2015Natur.522..455C}, [2] \cite{2017ApJ...851...40L}, [3] \cite{2015MNRAS.450.1846S}, [4] \cite{2018ApJ...854L...7C}, [5]  \cite{2017MNRAS.472..772M}, [6] \cite{2018MNRAS.481.1976Z}, [7]  \cite{Hutchison_2019}, [8] \cite{2015MNRAS.451.2050Z}, [9] \cite{2018MNRAS.477.2817S} [10] \cite{2017ApJ...851..145M}, [11] \cite{2018Natur.553..178S}, [12] \cite{2016MNRAS.462L...6K}, [13] \cite{2015A&A...574A..19S}, [14] \cite{2017ApJ...836L...2B}, [15] \cite{2020A&A...643A...2B}, [16] \cite{2016ApJ...829L..11P},  [17] \cite{2019ApJ...881..124M}. } 
\end{table}

\subsection{New VLT/X-SHOOTER observations}

We opted for deep X-SHOOTER observations with the purpose of investigating the \ion{C}{III}] emission at $\lambda1909 \ \AA$ in three star-forming galaxies at $z >5,$ having [\ion{C}{II}] detection at the location of the rest-frame UV emission (\citealt{2018MNRAS.478.1170C}). The three targets are scattered by about 1.5 dex around the local $\rm{L_{[\ion{C}{II}]}-SFR}$ relation, thus allowing us to probe the \ion{C}{III}] luminosity over a wide range of properties: $3\ \rm{M_{\odot} yr^{-1} \lesssim SFR \lesssim 100 \ M_{\odot} yr^{-1}}$, $10^7 L_{\odot} \lesssim L_{[\ion{C}{II}]} \lesssim 2.5 \times 10^9 L_{\odot}$, and $-4 \lesssim \log{(\rm{U})} \lesssim -1.5$. 
Additionally, the availability of multi-bands observations ({\it HST}, {\it Spitzer}, and ALMA) for these targets enabled us to correct the \ion{C}{III}] luminosity measurements from slit-losses (taking into account the size of the galaxy, the slit width, and the seeing) and dust extinction. 

In particular, HZ1 and HZ10 have been observed with ALMA in [\ion{C}{II}] by \cite{2015Natur.522..455C}, as part of their sample of normal star-forming galaxies at $5<z<6$.  Furthermore, the ALMA data has been reanalyzed by \cite{2018MNRAS.478.1170C}, with a focus on a detailed morphological analysis of the [\ion{C}{II}] emission. Specifically, for HZ10, the [\ion{C}{II}] line reveals multiple components and the velocity gradient implies a merger scenario. \cite{2018MNRAS.478.1170C} measured [\ion{C}{II}] luminosity of $L_{\ion{[C}{II}]} = 2.5 \pm 0.5 \times 10^8 L_{\odot}$ and $L_{\ion{[C}{II}]} = 25.1 \pm 5.0 \times 10^8 L_{\odot}$ for HZ1 and HZ10, respectively. \cite{2015Natur.522..455C} found the total SFR of HZ1 and HZ10 of $\rm{SFR}=24_{-3}^{+6} \ M_{\odot} \ \rm{yr^{-1}}$ and $\rm{SFR} = 169_{-27}^{+32} \ M_{\odot} \ \rm{yr^{-1}} $, respectively. Our third target, RXJ 1347-1216 is a Ly$\alpha$-emitting (LAE) galaxy at $z \sim 6.77$ lensed by the RXJ1347.1-1145 cluster (\citealt{2016ApJ...817...11H}).  \cite{2017ApJ...836L...2B} detected the  [\ion{C}{II}] line emission with ALMA and measured a luminosity of $L_{[\ion{C}{II}]} = 1.5_{-0.4}^{+0.2} \times 10^7 L_{\odot}$, and estimated a $\rm{SFR}=3.2 \pm 0.4 \ M_{\odot} \ \rm{yr^{-1}}$. These authors found that the galaxy lies below the local [\ion{C}{II}] - SFR relation (\citealt{2014A&A...568A..62D}), in agreement with the expectations from the \cite{2015ApJ...813...36V} model for a low-metallicity galaxy at $z \sim 7$. 

We observed our targets with the VLT/X-SHOOTER medium-resolution spectrograph  (\citealt{2011A&A...536A.105V}) on 02 April 2019, 19 April 2019, and 20 May 2019, respectively (European Southern Observatory (ESO) program ID: 0103.A-0692(A); PI: S. Carniani). The data for HZ1 and HZ10 were obtained over three observing blocks (OBs) of one hour each for every one of the three spectroscopic arms (UV/Blue (UVB), VISible (VIS), and Near-InfraRed (NIR)), in NODDING mode. An additional OB for each arm in STARE mode was carried out for each target, but it was not used in this work. For RXJ 1347-1216 source, only 1 OB was successfully obtained in NODDING mode for the VIS and NIR arms. The seeing variations were $0.31\arcsec-0.79\arcsec$ and $0.34\arcsec-0.86\arcsec$ for all OBs in the VIS and NIR arms, respectively. Sources were observed  under clear conditions. All OBs used a $1.2 \arcsec$ slit with a resolution of $R = 6500$ and $R = 4300$ for the VIS and NIR arms, respectively.

\subsection{Literature and archive sample}

 We included 11 sources at $6.0 < z < 7.5$ with \ion{C}{III}] and [\ion{C}{II}] emission from the literature (see Table \ref{detection} for references) to build a large sample of high-$z$ galaxies for our analysis. Out of these 11 systems, detection of both \ion{C}{III}] and [\ion{C}{II}] have been reported in two sources: COS-3018555981 (COS-3018 hereafter; \citealt{2017ApJ...851...40L}; \citealt{2018Natur.553..178S}) and A383-5.2 
 i.e., A383-5.1 (A383-5.2 hereafter\footnote{A383-5.1 and A383-5.2 are two images of the same galaxy lensed by the foreground A383 cluster. A383-5.2 was chosen as brighter of the two images in the rest-frame UV and it was detected in \ion{C}{III}], by \citealt{2015MNRAS.450.1846S}, whereas A383-5.1 was detected in [\ion{C}{II}] by \citealt{2016MNRAS.462L...6K}.}; \citealt{2015MNRAS.450.1846S}; \citealt{2016MNRAS.462L...6K}).

We also included GDS3073 and UVISTA-Z-002 with available X-SHOOTER archival data of the \ion{C}{III}] line and the [\ion{C}{II}] line data either published in the literature (for GDS3073; \citealt{2020A&A...643A...2B}) or available in the ALMA archive (for UVISTA-Z-002; \citealt{2021arXiv210613719B}). GDS3073  is a compact ($r \approx 0.11 \ \rm{kpc}$; \citealt{2010A&A...513A..20V}), active galactic nuclei (AGN) host galaxy at $z \sim 5.56$, with a possible ongoing merger (\citealt{2020ApJ...897...94G}) that has been detected in [\ion{C}{II}] by ALMA (labeled as  CANDELS\_GOODSS\_14; \cite{2020A&A...643A...2B} as part of the ALPINE survey ({\it ALMA Large Program to INvestigate C+ at Early times}; \citealt{2020A&A...643A...1L}). We retrieved and re-reduced (Sect. \ref{reduction_arch}) the  X-SHOOTER archival data for GDS3073 from two ESO programs: 384.A-0886 (PI: A. Raiter) and 089.A-0679 (PI: N. Norgaard), previously presented by \cite{2020ApJ...897...94G}. UVISTA-Z-002 galaxy at $z \sim 6.75$ was observed as a part of the REBELS (reionization-era bright emission line survey) ALMA large survey of $z>6.5$ luminous star-forming galaxies (\citealt{2021arXiv210613719B}). We retrieved and re-reduced (Sect. \ref{reduction_arch}) the X-SHOOTER archival data for UVISTA-Z-002 (ESO Program ID: 097.A-0082(A); PI:  R. Bowler). For this source, we also used the [\ion{C}{II}] observations from the ALMA archive in band 6 during Cycle 7 (Program ID: 2019.1.01634.L; PI: R. Bouwens).

Finally, we analyzed the VLT/MUSE (MultiUnit Spectroscopic Explorer) archival data (Program IDs: 094.A-0289(B), 095.A-0010(A), 096.A-0045(A), 099.A-0060(A), 098.A-0017(A), 096.A-0090(A), 0100.D-0807(C), 095.A-0240(A)) for a subsample of six galaxies at $1.75<z<1.95$ for which [\ion{C}{II}] detections were reported in \cite{2018MNRAS.481.1976Z}.

\section{Data reduction}\label{reduction}

\subsection{Data reduction of the new VLT/X-SHOOTER observations} \label{new-reduction}

\begin{figure*}
\centering
\includegraphics[width=\hsize]{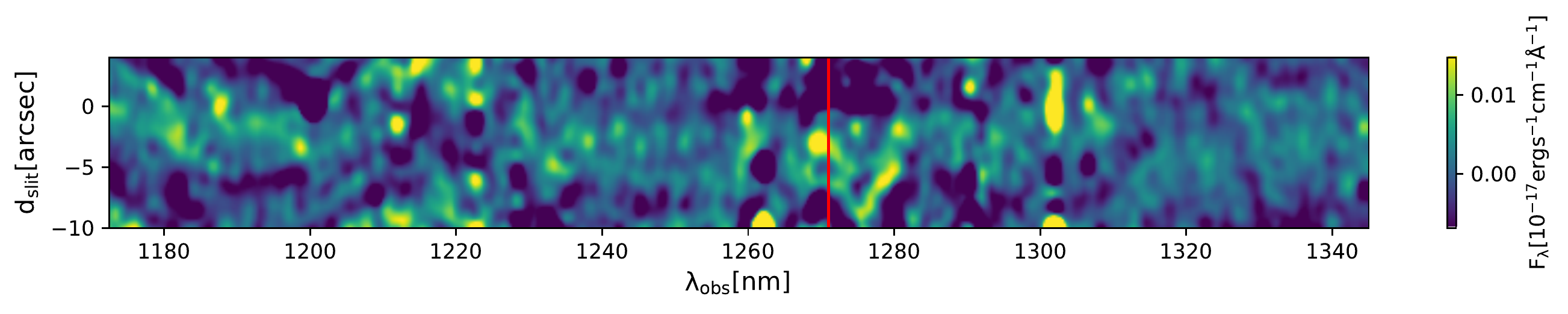}
\includegraphics[width=\hsize]{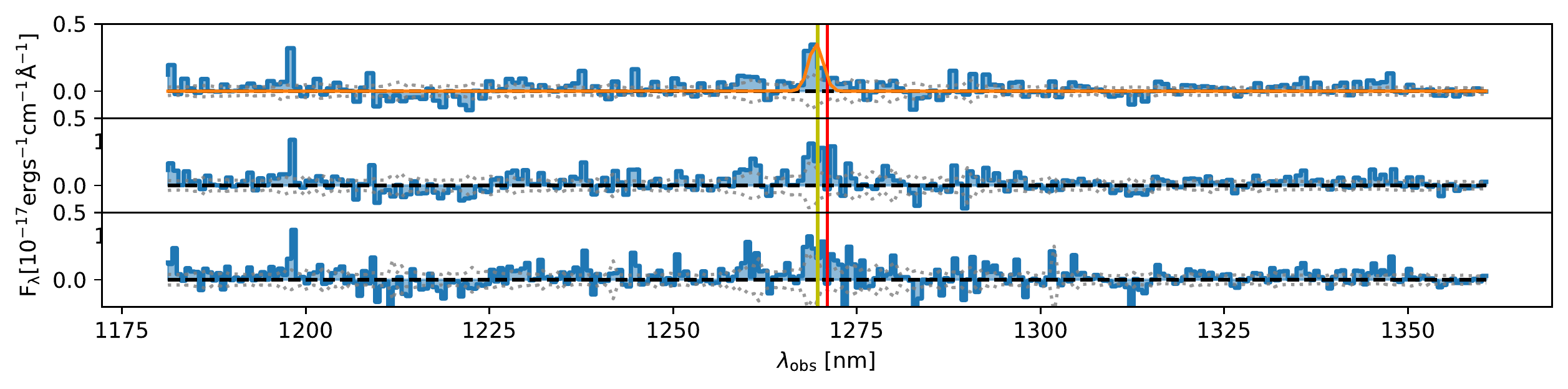}
 \caption{X-SHOOTER  NIR spectrum of the HZ10 galaxy. Top panel shows the combined, rebinned ($\approx 85 \ \rm{km \ s^{-1}}$), and smoothed ($\rm{1.1 \ nm \times 0.12 \arcsec}$) 2D spectrum. The \ion{C}{III}] $\lambda \lambda1907, 1909$ line is detected in the center of the slit with the negative flux on each side. The bottom three panels illustrate the extracted 1D spectra rebinned by $\approx 210\ \rm{km \ s^{-1}}$,  $\approx 170\ \rm{km \ s^{-1}}$, and $\approx 140\ \rm{km \ s^{-1}}$, respectively. Shaded blue regions indicate the flux, dotted lines indicate the uncertainties, whereas orange solid line is the best Gaussian fit. Expected positions of the \ion{[C}{III]} $\lambda 1907$ at  $\lambda = 1269.7$ nm and \ion{C}{III]} $\lambda 1909$ line at  $\lambda = 1271.0$ nm, based on the Ly$\alpha$ spectroscopic redshift, are shown as yellow and red lines, respectively. 
       }
     \label{2D-HZ10}
\end{figure*}

 We adopted the X-SHOOTER pipeline (\citealt{Modigliani}) in the REFLEX (REcipe FLexible EXecution)  environment\footnote{\url{https://www.eso.org/sci/software/esoreflex/}} to reduce the raw  data. We also developed a \texttt{Python} script for the data analysis on the ESO pipeline final data products. We stacked the individual 2D spectra extracted from the OBs of each target, excluding bad pixels, and obtained a final weighted mean 2D spectra for the VIS and NIR spectra separately, by using the Drizzle package.\footnote{\url{https://github.com/spacetelescope/drizzle}} Next, we cropped the output 2D spectra at the wavelength range and slit position where we expect to see the Ly$\alpha$ and \ion{C}{III}] lines in the VIS and NIR spectra, respectively. In order to increase the signal-to-noise ratio (S/N) of the emission lines, we also performed different spectral rebinning, depending on the expected width of the lines.
 Stacked, rebinned, and smoothed X-SHOOTER 2D NIR spectra, and the extracted 1D spectra for HZ10, HZ1, and RXJ 1347-1216  are shown in  Figs. \ref{2D-HZ10}, \ref{2D-HZ1}, and \ref{2D-RXJ},  respectively. 
 
\subsection{Reduction of the archival data} \label{reduction_arch}

 The retrieved X-SHOOTER data of GDS3073 contains 21 OBs, of which 11 OBs from the program ID 384.A-0886 and 10 OBs from the program ID: 089.A-0679, for each arm. The NIR observations of the two programs are obtained in the different wavelength range. Therefore, we first  stacked the individual 2D NIR spectra of each program separately, then we cropped the two 2D NIR spectra to be in the same wavelength range. We obtained the final 2D NIR spectrum as the weighted mean of the two 2D NIR spectra of each program.  In order to extract the final 1D spectra, we followed the same procedure as described in Subsection \ref{new-reduction}. The results are shown in  Fig. \ref{2D-GDS}.
 
 The X-SHOOTER archival data of UVISTA-Z-002 were retrieved with four OBs, for each of the three arms (ESO Program ID: 097.A-0082(A); PI:  R. Bowler). For the data reduction step, we repeated the steps as in Subsection \ref{new-reduction}.  The results are shown in Fig. \ref{2D-UVISTA}.
 For this source, we also reduced ALMA observations of the [\ion{C}{II}] line by using CASA (common astronomy software applications) software\footnote{\url{https://casa.nrao.edu/}} (\citealt{2007ASPC..376..127M}) and running the \texttt{Python} script delivered with the observations. 

Finally,  we used the VLT/MUSE archival data for a subsample of $1.75<z<1.95$ galaxies for which [\ion{C}{II}] line detection has been reported in \cite{2018MNRAS.481.1976Z}. We extracted the 1D spectra from a circular region with a radius of $r =5\arcsec$, enclosing the position of the target galaxy, and we analyzed the results using ds9 software\footnote{\url{https://ds9.si.edu}} and a \texttt{Python} script we developed. However, we still did not detect any \ion{C}{III}] line emission in the extracted 1D spectrum of any source or in the stacked 1D spectrum of all the six galaxies from the MUSE subsample (see Fig. \ref{1D-MUSE}).

%

\section{X-SHOOTER data analysis: \ion{C}{III}] line emission} {\label{xshooter_analysis}}

We visually inspected the X-SHOOTER VIS and NIR 1D spectra  for the Ly$\alpha$, \ion{C}{III}], and any additional rest-frame UV emission line for each source. Next, we rebinned the 1D spectra to increase the signal-to-noise ratio and we fitted the line emission with a Gaussian to derive the integrated flux.

We recovered the Ly$\alpha$ emission detection in HZ1 and HZ10, previously reported in \cite{2015Natur.522..455C}  with the DEIMOS spectrograph on the Keck II telescope. We also identified a hint of the Ly$\alpha$ emission in RXJ 1347-1216, previously published in \cite{2016ApJ...817...11H} with DEIMOS. Next, in the NIR 2D spectrum and the extracted 1D spectra of the HZ10 galaxy, we identified the blended \ion{C}{III}] $\lambda \lambda1907, 1909$ doublet line at the expected redshifted wavelength position of the \ion{C}{III}] line, at  $\lambda \approx 1270 $ nm (Fig. \ref{2D-HZ10}, top panel). The feature with a positive flux is in the center of the slit with two negative flux features on each side of the positive emission (Fig. \ref{2D-HZ10}, top panel), as data were taken in the NODDING mode. 

The X-SHOOTER spectrograph allows us to resolve the \ion{C}{III}] $\lambda \lambda1907, 1909$ doublet. However, in order to detect the \ion{C}{III}] with a statistically significant S/N, we need to rebin the spectra via 15 channels ($\approx 200\ \rm{km \ s^{-1}}$), which decreases the spectral resolution. Therefore, we did  not ultimately detect the two lines separately,  so we fit the \ion{C}{III}] doublet  with a single Gaussian, from which we obtained a total integrated flux of $\rm{8.8 \pm 2.1 \times 10^{-17} \ erg \ s^{-1} \ cm^{-2}}$ with a S/N = 4.2 and a $\rm{FWHM = 550.9 \pm 98.8 \ km\ s^{-1}} $. The peak of the doublet line occurs at $\lambda = 1269.4 \pm 0.2 $ nm, which is slightly blueshifted than the expected positions of the [\ion{C}{III}] $\lambda1907$ line  at $\lambda = 1269.7$ nm and \ion{C}{III}] $\lambda1909$ line  at $\lambda = 1271.0 $ nm, based on the Ly$\alpha$ spectroscopic redshift from \cite{2015Natur.522..455C}  (Fig. \ref{2D-HZ10}, yellow and red vertical lines, respectively). However, at high-$z$, a significant part of the Ly$\alpha$ emission can be absorbed by the increasing fraction of the neutral medium and the peak of the observed Ly$\alpha$ emission might be offset with respect to the systematic redshift (\citealt{2014PASA...31...40D}). If we take the spectroscopic redshift based on the [\ion{C}{II}] line emission from \cite{2015Natur.522..455C}, the expected positions of the two \ion{C}{III}] lines are at $\lambda = 1269.2 $ nm and $\lambda = 1270.5 $ nm,  and the peak of the doublet line occurs between the two separate lines. 

Assuming a $[\ion{C}{III}]  \ \lambda1907/\ion{C}{III}] \ \lambda1909$ line ratio of $1.4 \pm 0.2$ (\citealt{2015MNRAS.450.1846S}), we estimated the $\ion{C}{III}]\ \lambda1909$ line integrated flux of $\rm{3.7 \pm 1.1 \times 10^{-17} \ erg \ s^{-1} \ cm^{-2}}$ with S/N = 3.4. For the other two targets, HZ1 and RXJ 1347-1216, we did not detect \ion{C}{III}] emission at the expected redshifted wavelength and we place a $3\sigma$ upper limit on the line flux. For HZ1 and RXJ 1347-1216, we estimate the integrated flux upper limit assuming a full-width at half maximum (FWHM) of the \ion{C}{III}]  line to be equal to the FWHM of the [\ion{C}{II}] line, where $\rm{FWHM_{[\ion{C}{II}]} = 72 \pm 11 \ km \ s^{-1}}$ (\citealt{2015Natur.522..455C}) and $\rm{FWHM_{[\ion{C}{II}]} = 75 \pm 25 \ km \ s^{-1}}$  (\citealt{2017ApJ...836L...2B}), respectively. Furthermore, we inspected the VIS and NIR spectra of our targets, but we did not detect any additional UV-rest frame lines.

 GDS3073 has been observed with FORS2, VIMOS, and X-SHOOTER spectrographs at the VLT telescope and several UV rest-frame line detection have been reported, including  $\ion{N}{V}] \ \lambda1240$ and $\ion{O}{VI} \ \lambda1032-1038$, confirming the AGN nature of GDS3073 (\citealt{2010A&A...513A..20V}; \citealt{2020ApJ...897...94G}). We recover a very bright Ly$\alpha$ emission detection with a S/N $\sim 22$ and \ion{N}{IV]} line emission with a S/N $\sim 3.9$. For the \ion{C}{III}] line, we place an upper limit on the  velocity integrated flux, assuming  $\rm{FWHM_{\ion{C}{III}]} = FWHM_{[\ion{C}{II}]}} $, where the $\rm{FWHM_{[\ion{C}{II}]} = 230 \pm 16 \ km \ s^{-1}}$ derived by \cite{2020A&A...643A...2B}. In the spectra of UVISTA-Z-002, we did not detect any emission line. We placed a $3\sigma$ upper limit on the \ion{C}{III}] line flux and estimated the velocity integrated flux upper limit, assuming  $\rm{FWHM_{\ion{C}{III}]} = FWHM_{[\ion{C}{II}]}} $, where  $\rm{FWHM_{[\ion{C}{II}]}} = 45.2 \pm 7.6 \rm{\ km \ s^{-1}}$ was obtained from the Gaussian fit on the observed [\ion{C}{II}] line from this work. 
 
 For the intermediate-$z$ galaxy sample, we did not identify any emission at the expected location of the \ion{C}{III}] line in the extracted 1D spectra of each individual galaxy nor in the stacked spectrum. Therefore, we set an upper limit to the \ion{C}{III}] flux and we estimated the velocity integrated flux assuming  $\rm{FWHM_{\ion{C}{III}]} = FWHM_{[\ion{C}{II}]}} $, with the FWHM of the [\ion{C}{II}] line estimates from \cite{2018MNRAS.481.1976Z}. The observed \ion{C}{III}] line flux of the entire sample of galaxies is in Table \ref{CIII-prop}.

 \begin{table*}[h]
 \caption[]{\label{CIII-prop} UV radius, observed and intrinsic \ion{C}{III}] line emission flux, \ion{C}{III}] intrinsic luminosity, and surface brightness of the full galaxy sample.}
\begin{tabular}{lcccccccc}
 \hline \hline
  ID & $r_{\rm{UV}}$ &  $F_{\rm{{\ion{C}{III}]}}}^{\rm{obs}}$ & $F_{\rm{{\ion{C}{III}]}}}^{\rm{int}}$ & $L_{\rm{\ion{C}{III}]}}^{\rm{int}}$ &  $\Sigma_{\rm{\ion{C}{III}]}}^{\rm{int}}$
  &
  Ref. \\
 & (kpc) & ($\rm{10^{-18} erg \ s^{-1}  cm^{-2}}$)  &($\rm{10^{-18} erg \ s^{-1}  cm^{-2}}$) & ($10^8 L_{\odot}$) &  ($10^8 L_{\odot} \ \rm{kpc^{-2}}$) &
 \\ \hline
 \noalign{\smallskip}
 \multicolumn{7}{c}{New \ion{C}{III}] data on galaxies at $5.5 < z< 7$} \\
 \noalign{\smallskip}
 \hline
 \noalign{\smallskip}
HZ10 & $1.1^a$ & $36.7 \pm 11.1$ & $64.7_{-19.8}^{+20.1}$ & $61.5_{-18.8}^{+19.1} $ & $8.1 \pm 4.1 $ & [1, $\dag$]\\
 \noalign{\smallskip}
HZ1 & 1.2 & <8.7 &<8.7 &<8.4 & <0.9 & [1, $\dag$] \\
 \noalign{\smallskip}
RXJ 1347-1216 & 0.8 & <5.2 & <5.2 & <7.5 & <1.8 & [2, $\dag$]\\
 \noalign{\smallskip}
GDS3073  & $0.11 \pm 0.01$ &<2.6 &<2.6 & <2.4 & <31.6 & [3, $\dag$]\\
 \noalign{\smallskip}
UVISTA-Z-002  &  $2.1 \pm 0.1^b$ &<1.5 &<1.5 & <2.0  & <0.3 & [$\dag$]\\
 \noalign{\smallskip}
\hline
 \noalign{\smallskip}
 \multicolumn{7}{c}{New \ion{C}{III}] data on galaxies at $z \sim 2$} \\
\noalign{\smallskip}
\hline
 \noalign{\smallskip}
9347  &$3.08 \pm 0.07$  &<336.2 &<336.2 & <22.1 & <0.4 & [$\dag$]\\ %
 \noalign{\smallskip}
6515 &$2.2 \pm 0.4$ &<605.9 &<605.9& <39.4 & <1.3 & [$\dag$]\\ %
 \noalign{\smallskip}
10076 & $2.35 \pm 0.03$  & <1430.4 &<1430.4 & <106.2 & <3.1 & [$\dag$] \\ %
 \noalign{\smallskip}
9834 &$0.97 \pm 0.02$ &<2115.6 &<2115.6 & <123.6 & <20.9 & [$\dag$]\\ %
 \noalign{\smallskip}
9681  &$1.93 \pm 0.03$ & <1088.9 &<1088.9& <74.8 & <3.2 & [$\dag$]\\ %
 \noalign{\smallskip}
8490 &$1.6 \pm 0.01$&<536.0 &<536.0 & <37.8 & < 2.4 & [$\dag$]\\ %
 \noalign{\smallskip}
\hline
\noalign{\smallskip}
 \multicolumn{7}{c}{Literature sample of galaxies at $6 < z < 7.5$} \\
\noalign{\smallskip}
\hline
 \noalign{\smallskip}
COS-3018 &$1.3 \pm 0.1^b$ & $1.3 \pm 0.3$ & $8.5 \pm 6.8$ &  $12.6 \pm 10.2$ & $1.3 \pm 1.0$ & [4, 6] \\
 \noalign{\smallskip}
A383-5.2  &$0.9$ & $3.7 \pm 1.1$ &$5.6 \pm 1.9$ & $6.2 \pm 2.1$ & $1.1 \pm 0.6$ & [1, 7]\\ %
 \noalign{\smallskip}
 z7\_GND\_42912 & $0.72 \pm 0.01$ &$2.6 \pm 0.5$ &$12.0 \pm 5.7$ & $2.2 \pm 1.1$ & $6.8 \pm 3.2$ & [\dag, 8]\\ %
 \noalign{\smallskip}
Himiko  & $0.9^a$  &<5.8 &<5.8 & <7.9 & <1.6 & [1, 9]\\ %
 \noalign{\smallskip}
CR7  &$0.9^a$  &<17.0 &<17.0& <23.1 & <4.6 & [1, 10]\\ %
 \noalign{\smallskip}
CR7a  &0.9  &<15.0 &<15.0 & <20.4& <4.0 & [1, 10]\\ %
 \noalign{\smallskip}
CR7b  &0.9  & <13.0 &<13.0 & <18.0 & <3.5 & [1, 10]\\  %
 \noalign{\smallskip}
CR7c &0.9  &<14.0 &<14.0 & <19.1 & <3.7 & [1, 10]\\ %
 \noalign{\smallskip}
COS-13679 & 0.8  &<0.8 &<0.8 & <1.4& <0.4 & [1, 6]\\ %
 \noalign{\smallskip}
COS-2987 &$1.6 \pm 0.5^b$  & <1.6 &<1.6 & <2.3 & <0.2 & [4, 6]\\ %
 \noalign{\smallskip}
VR7 &$1.6 \pm 0.1$  & <21.0&<21.0 &<27.9 & <1.8 & [5, 11]\\ %
 \noalign{\smallskip}
\hline
\end{tabular}
\tablefoot{We assume 20\% uncertainties on the $r_{\rm{UV}}$ when they are not provided. $^a$ UV radius of the largest component of a multi-component system. $^b$ UV radius is estimated to be one half of the [\ion{C}{II}] radius.}
\tablebib{Redshift, $r_{\rm{UV}}$, and \ion{C}{III}] flux references: [$\dag$] This work, [1] \cite{2018MNRAS.478.1170C}, [2] \cite{2017ApJ...836L...2B},  [3] \cite{2010A&A...513A..20V}, [4] \cite{2018Natur.553..178S}, [5] \cite{2019ApJ...881..124M},  [6] \cite{2017ApJ...851...40L},  [7] \cite{2015MNRAS.450.1846S}, [8] \cite{Hutchison_2019}, [9] \cite{2015MNRAS.451.2050Z}, [10] \cite{2018MNRAS.477.2817S}, [11] \cite{2017MNRAS.472..772M}. 
}
\end{table*}

\section{Results} \label{Results}

In the previous section, we derived the $\ion{C}{III}]\ \lambda1909$ line integrated flux of $F_{\ion{C}{III}]}^{\rm{obs}} = \rm{3.7 \pm 1.1 \times 10^{-17} \ erg \ s^{-1} \ cm^{-2}}$ for HZ10. We also included a literature sample of three $z>6$ galaxies: COS-3018 (\citealt{2017ApJ...851...40L}; \citealt{2018Natur.553..178S}), A383-5.2 (\citealt{2015MNRAS.450.1846S}; \citealt{2016MNRAS.462L...6K}), and z7\_GND\_42912 (\citealt{Hutchison_2019}; \citealt{2015A&A...574A..19S}),  with the observed \ion{C}{III}] integrated flux in the range of $F_{\ion{C}{III}]}^{\rm{obs}} = \rm{1.3-3.7 \times 10^{-18} \ erg \ s^{-1} \ cm^{-2}}$.

As a rest-frame UV line at $\lambda = 1909 \ \AA$, the $\ion{C}{III}]$ is expected to be affected by dust attenuation (see, e.g., \citealt{2020ARA&A..58..529S} and references therein). Therefore,  we correct the \ion{C}{III}] emission for the dust attenuation and derive the attenuation-corrected \ion{C}{III}] integrated flux for galaxies with \ion{C}{III}] detection, whereas for non-detections, we did not perform this correction. Next, we constrained the burstiness parameter, $\kappa_s$, metallicity, $Z$, and the total density of the emitting gas, $n$, with the use of the  \cite{2020MNRAS.495L..22V} model. Finally, we compare the ISM properties of HZ10 and COS-3018 derived in this work and in the literature. 

\subsection{Intrinsic \ion{C}{III}] line emission}

  We derived the UV attenuation at $1900 \ \AA$, following \cite{1999ApJ...521...64M}:

\begin{equation} \label{schouws}
    A_{1900} = \frac{dA_{\rm{UV}}}{d\beta_{\rm{UV}}} \ (\beta_{\rm{UV}} - \beta_{\rm{UV,  intr}}),
\end{equation}%
where $\beta_{\rm{UV}}$ is the UV-continuum slope ($f_{\lambda} \propto \lambda^{\beta_{\rm{UV}}}$; \citealt{1994ApJ...429..582C}) of sources with \ion{C}{III}] detection from the literature (Table \ref{slope}), ${dA_{\rm{UV}}}/{d\beta_{\rm{UV}}}$ is the steepness of the dust attenuation law and $\beta_{\rm{UV, intr}}$ is the intrinsic UV-continuum slope. \cite{1999ApJ...521...64M} has derived this correlation to estimate the UV attenuation at 1600 $\AA$. However, we use the same correlation to derive the UV attenuation at 1900 $\AA$, assuming that the attenuation is nearly identical at two wavelengths. Next, to make use of  Eq. (\ref{schouws}), we need to determine the ${dA_{\rm{UV}}}/{d\beta_{\rm{UV}}}$ and $\beta_{\rm{UV, intr}}$ parameters, which depend on the dust attenuation curve of a galaxy. In order to constrain the dust attenuation law, we can use the $\rm{IRX}-\beta_{\rm{UV}}$ diagram, where IRX is the infrared excess parameter $\rm{IRX = L_{FIR}/L_{UV}}$. HZ10 is a massive, metal-rich, star-forming galaxy (\citealt{2015Natur.522..455C}; \citealt{2017ApJ...847...21F}) that falls on the local starburst relation (\citealt{2017ApJ...845...41B}; \citealt{2017ApJ...847...21F}), with  ${dA_{\rm{UV}}}/{d\beta_{\rm{UV}}} = 1.99$ and $\beta_{\rm{UV, intr}} = -2.23$ (\citealt{1999ApJ...521...64M}). However, for the remaining three sources: COS-3018, A383-5.2 and z7\_GND\_42912, there are only upper limits on the $L_{\rm{FIR}}$ and, thus, the IRX parameter, are reported in the literature, which also depends on the assumption on the dust temperature (\citealt{2018MNRAS.477..552B}; \citealt{2020MNRAS.497..956S, 2022MNRAS.513.3122S}). Therefore, it is not possible to constrain the dust attenuation curve of these objects, since they could fall on any of the many variants of the dust attenuation law (e.g., \citealt{1999ApJ...521...64M}). Therefore, for these $z>5$ sources, we estimated the UV attenuation, $A_{1900}$, assuming that they lie on the fiducial relation for local starbursts, in agreement with results for high-$z$ galaxies from the literature  (\citealt{2020MNRAS.491.4724F}; \citealt{2021MNRAS.501.3289V}). Values of the derived UV attenuation $A_{1900}$ are given in Table \ref{slope}.

\begin{table}[h]
 \caption[]{\label{slope}   $\beta_{\rm{UV}}$ slope and UV attenuation at $1900 \AA$ for three $z> 5$ galaxies with \ion{C}{III}] and [\ion{C}{II}] detections.}
\begin{tabular}{lcccccccc}
 \hline \hline
  \noalign{\smallskip}
  ID & $\beta_{\rm{UV}}$ & $A_{1900}$& Ref. \\
   \noalign{\smallskip}
 \hline
 \noalign{\smallskip}
HZ10 & $-1.92_{-0.17}^{+0.24} $& $0.6 \pm 0.1$ & [1]\\
 \noalign{\smallskip}
COS-3018 & $-1.22 \pm 0.51$& $2.0 \pm 0.8$ & [2] \\
 \noalign{\smallskip}
A383-5.2  & $-2 \pm 0.7$& $0.5 \pm 0.2$ & [3]\\ %
 \noalign{\smallskip}
 z7\_GND\_42912 & $-1.4 \pm 0.4$ & $1.7 \pm 0.5$ & [4]\\ %
 \noalign{\smallskip}
\hline
\end{tabular}
\tablebib{$\beta_{\rm{UV}}$ slope references:  [1] \cite{2017ApJ...845...41B}, [2] \cite{2018Natur.553..178S}, [3] \cite{2019ApJ...881..124M}, [4] \cite{2015A&A...574A..19S}.}
\end{table}

The absorption law slope for local starbursts found by \cite{1999ApJ...521...64M} is consistent with the Calzetti dust attenuation law (\citealt{2000ApJ...533..682C}). Thus, the intrinsic \ion{C}{III}] flux of high-$z$ sources can be derived from the observed \ion{C}{III}] flux using the Calzetti attenuation law as 

\begin{equation}
    F_{\rm{{\ion{C}{III}]}}}^{\rm{int}} = F_{\rm{{\ion{C}{III}]}}}^{\rm{obs}} \  10^{0.4 A_{1900}}.
\end{equation}

For HZ10, we derived the intrinsic flux of the \ion{C}{III}] line of $\rm{6.5 \pm 2.0 \times 10^{-17} \ erg \ s^{-1} \ cm^{-2}}$, which is higher than the observed value by a factor
of $\approx 1.8$. 

We derived the intrinsic surface brightness of the $\ion{C}{III}]$ line as:

\begin{equation} \label{sigma}
\Sigma_{\ion{C}{III}]}^{\rm{int}} = \frac{L_{\ion{C}{III}]}^{\rm{int}}}{2 \pi r_{\rm{UV}}^2},
\end{equation}%
where $\rm{r_{UV}}$ is the UV radius in kpc tracing the \ion{C}{III}] emitting ionized phase of the ISM, $L_{\ion{C}{III}]}^{\rm{int}} = 4 \pi D_L^2 F_{\ion{C}{III}]}$ is the intrinsic luminosity in solar luminosity $L_{\odot}$ and $D_L$ is the luminosity distance in Mpc. For the z7\_GND\_42912 source, we derive the UV radius $ r_{\rm{UV}}$  using the {\it Hubble Space Telescope} ({\it HST}) Wide-Field Camera 3 (WFC3) image of the galaxy in the F160W filter, as well as the QFitsView\footnote{\url{https://www.mpe.mpg.de/~ott/QFitsView/}} and Aladin\footnote{\url{https://aladin.u-strasbg.fr/}} astronomical tools. We fit the image of the galaxy with an elliptical Gaussian and measure the FWHM in pixels. Next, we derived the FWHM in $\arcsec$, deconvolved by the FWHM of the  point-spread function (PSF) for the ({\it HST}) WFC3 camera of $\rm{FWHM}_{\rm{PSF}} = 0.18 \pm 0.01 \ arcsec$ (e.g., \citealt{2012ApJ...745...85L}). Finally, we derived the deconvolved UV radius in kpc. For most of the galaxies included in our analysis, we use $ r_{\rm{UV}}$ from the literature. For UVISTA-Z-002, COS-3018, and COS-2987, $ r_{\rm{UV}}$ is not measured in the literature, so we assume $r_{\rm{UV}} \sim 0.5 \ r_{\rm{\ion{[C}{II}]}}$, following \cite{2018MNRAS.478.1170C}. For six intermediate-$z$ galaxies, we estimate the UV radii $r_{\rm{UV}}$  using the {\it HST} Advanced Camera for Surveys (ACS) images in the $b$ magnitude, with a  $\rm{FWHM}_{\rm{PSF}} = 0.08 \ arcsec$ (e.g., \citealt{2011ApJS..193...27W}) and following the above-mentioned procedure. 
The spectroscopic redshift, UV radius, observed, and intrinsic \ion{C}{III}] line emission flux, \ion{C}{III}] intrinsic luminosity and \ion{C}{III}] intrinsic surface brightness of the entire galaxy sample used in this work are given in Table \ref{CIII-prop}.

\subsection{[\ion{C}{II}] line emission and SFR}

For UVISTA-Z-002, we fit the observed [\ion{C}{II}] line emission with a Gaussian from which we derived an integrated flux of $35.5 \pm 7.7 \ \rm{mJy \ beam^{-1} \ km \ s^{-1}}$, with a S/N $= 4.5$ and a FWHM$ = 45.2 \pm 7.6 \rm{\ km \ s^{-1}}$. Peak of the Gaussian is at $1204.1 \ \rm{\mu m}$, corresponding to a spectroscopic redshift of $z_{\rm{sp}} = 6.634 \pm 0.004$ which is consistent with the previous measurement of the photometric redshift $z_{\rm{ph}} = 6.75$, with uncertainties in the range of $\Delta z_{\rm{ph}} = 0.07-0.09$ (\citealt{2021arXiv210613719B}). Next, we derived the [\ion{C}{II}] surface brightness $\Sigma_{[\ion{C}{II}]}$ as:

\begin{equation}
\Sigma_{[\ion{C}{II}]} = \frac{L_{ [\ion{C}{II}]}}{2 \pi r_{\rm{[\ion{C}{II}]}}^2},
\end{equation}%
where $L_{ [\ion{C}{II}]} = 1.04 \times 10^{-3} S_{ [\ion{C}{II}]} \Delta v D_L^2$ is [\ion{C}{II}] luminosity in units of solar luminosity from \cite{2013ARA&A..51..105C}, $S_{ [\ion{C}{II}]} \Delta v$ is the velocity integrated flux of the [\ion{C}{II}] line, $D_L$ is the luminosity distance in Mpc, $\nu_{\rm{obs}}$ is the observed frequency in GHz, and $r_{ [\ion{C}{II}]}$ is the [\ion{C}{II}] radius in kpc. For UVISTA-Z-002, we measure $r_{[\ion{C}{II}]}$ from the ALMA continuum-subtracted, [\ion{C}{II}] line map, and the imfit task of the CASA software (version-6.4.0-16). We fit the  [\ion{C}{II}] emitting region of the source with an elliptical Gaussian and measure the beam-deconvolved FWHM in $\arcsec$, from which we derived the $r_{[\ion{C}{II}]}$ in kpc. For the other sources from our sample, we used the $r_{ [\ion{C}{II}]}$ values from the literature when available or we assume $r_{[\ion{C}{II}]} \sim 2r_{\rm{UV}}$ following \cite{2018MNRAS.478.1170C}.

\begin{figure}
\centering
\includegraphics[width=\hsize]{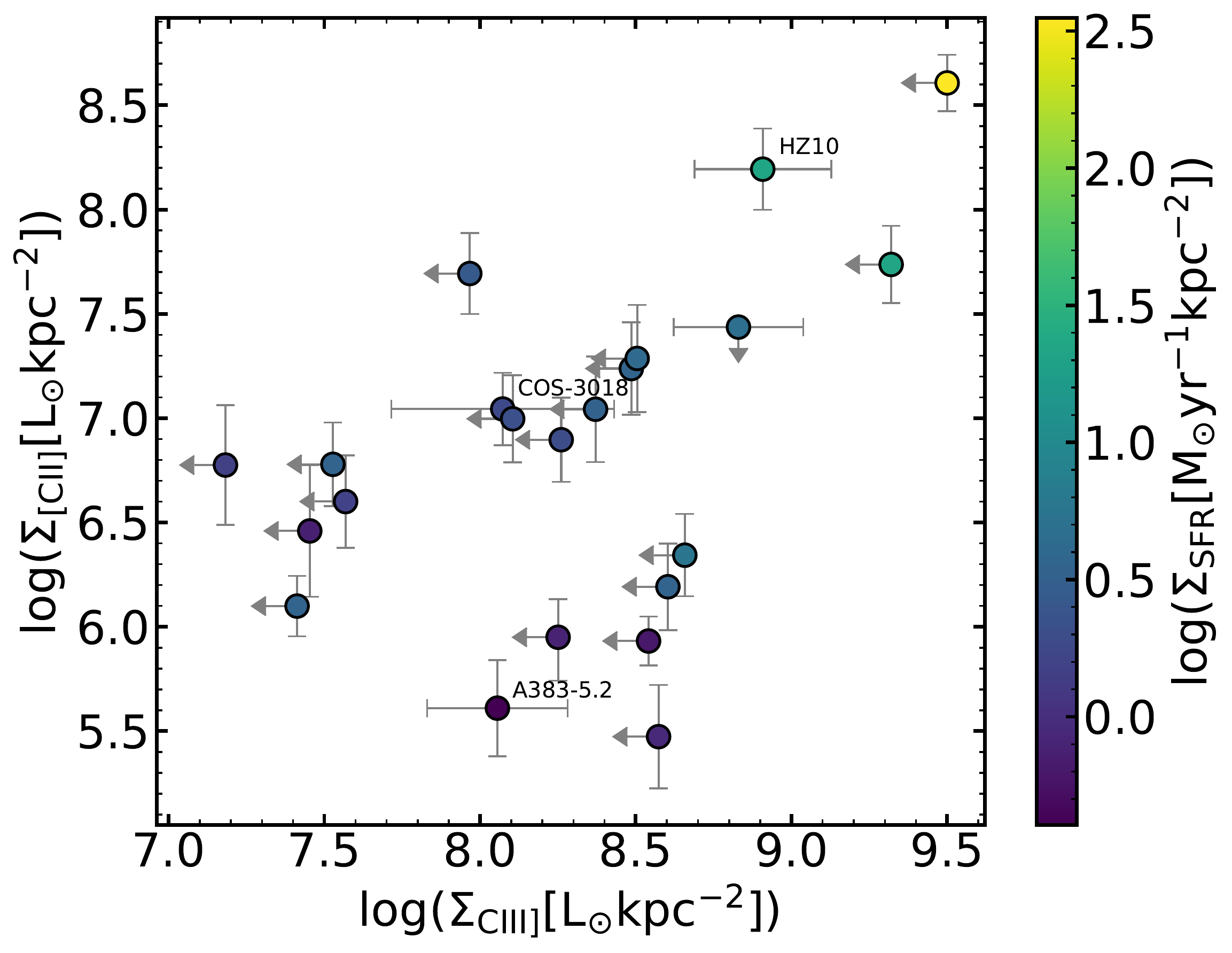}
 \caption{[\ion{C}{II}] surface brightness $\Sigma_{[\ion{C}{II}]}$ vs. \ion{C}{III}] surface brightness $\Sigma_{\ion{C}{III}]}$, color-coded by the SFR per unit area $\Sigma_{\rm{SFR}}$, for our entire sample of galaxies. Sources that are detected both in [\ion{C}{II}] and \ion{C}{III}] are annotated.  
          }
     \label{Full_sample}
\end{figure}

To constrain the ISM parameters with the \cite{2020MNRAS.495L..22V} method, we also included the total SFR per unit area, which is derived as:

\begin{equation}
\rm{\Sigma_{\rm{SFR}} = \frac{SFR}{2 \pi r_{\rm{UV}}^2}},
\end{equation}%
where $\rm{SFR = SFR_{UV}} + SFR_{IR}$ is the total SFR and $r_{\rm{UV}}$ is the UV radius tracing the extension of the star-forming regions (\citealt{2018MNRAS.478.1170C}). We either make use of the $\rm{SFR_{UV}}$ and $\rm{SFR_{IR}}$ given in the literature, converted to the \cite{2012ARA&A..50..531K} scaling relations, when necessary; or we constrain these parameters using the corresponding UV and IR luminosities, along with the scaling relations of \cite{2012ARA&A..50..531K}. For GDS3073 and the $z \sim 2$ sources of our sample, the UV luminosity $L_{\rm{UV}}$ is not reported in the literature. Therefore, for these objects we computed the $L_{\rm{UV}}$ from the absolute magnitudes corresponding to rest-frame UV (\citealt{2020MNRAS.494.1071G}).
Moreover, for some high-$z$ galaxies no $\rm{SFR_{IR}}$ estimates or only upper limits are provided in the literature. For those sources, we assumed $\rm{SFR = SFR_{UV}}$. Table \ref{CII-SFR} shows the [\ion{C}{II}] radius, luminosity and surface brightness, and total SFR and SFR per unit area for the full galaxy sample. Figure \ref{Full_sample} shows [\ion{C}{II}] surface brightness versus \ion{C}{III}] surface brightness, color-coded with the SFR per unit area, for our entire sample of galaxies. We observe that galaxies with high \ion{C}{III}] and [\ion{C}{II}] surface brightness tend to experience  high SFR per unit area, although the data are scattered and most of the sources have only \ion{C}{III}] upper limits.

\begin{table*}[h]
 \caption[]{\label{CII-SFR}[\ion{C}{II}] radius, luminosity and surface density, and total SFR, and SFR per unit area of the full galaxy sample.}
\begin{tabular}{lcccccccc}
 \hline \hline
  \noalign{\smallskip}
  ID & $r_{\rm{[\ion{C}{II}]}}$ & $L_{\rm{[\ion{C}{II}]}}$ & $\Sigma_{\rm{[\ion{C}{II}]}}$ 
 & SFR & $\Sigma_{\rm{SFR}}$ &
  Ref. \\
 &  (kpc)&  ($10^8 L_{\odot}$)& ($10^8 L_{\odot} \ \rm{kpc^{-2}}$) & ($M_{\odot} \ \rm{yr^{-1}} $) &($M_{\odot} \ \rm{yr^{-1} \ kpc^{-2}}$)& \\ 
  \noalign{\smallskip}
 \hline
 \noalign{\smallskip}
 \multicolumn{7}{c}{New \ion{C}{III}] data on galaxies at $5.5 < z< 7$} \\
 \noalign{\smallskip}
 \hline
 \noalign{\smallskip}
 HZ10 & 1.6 & $25.1 \pm 5.0 $ & $1.6 \pm 0.7 $ & $169_{-27}^{+32}$ &  $22.2_{-9.6}^{+9.8}$ & [1, 8] \\
 \noalign{\smallskip}
HZ1 &  0.9&$2.5 \pm 0.5 $ & $0.5 \pm 0.2 $ &$24_{-3}^{+6}$ &  $2.7_{-1.1}^{+1.2}$ & [1, 8]  \\
 \noalign{\smallskip}
RXJ 1347-1216 & $1.6^{a}$ & $0.15_{-0.04}^{+0.02}$& $0.009_{-0.004}^{+0.004} $ &  $3.2 \pm 0.4^c$ & $0.8 \pm 0.3$ &[2, 9] \\
 \noalign{\smallskip}
GDS3073 & $0.22 \pm 0.02^{a}$ & $1.2 \pm 0.3$& $4.1 \pm 1.3 $  &  $26.8 \pm 5.4^c$ & $352.7 \pm 95.3$ &  [3, 10, \dag] \\
 \noalign{\smallskip}
 UVISTA-Z-002 & $(4.2 \pm 0.1)\times(1.2 \pm 0.3)$ & $0.4 \pm 0.1 $ & $0.01 \pm 0.004 $ & $27.7^c$ &  $3.5 \pm 1.1$   & [\dag, 7] \\
\hline
\noalign{\smallskip}
 \multicolumn{7}{c}{ New \ion{C}{III}] data on galaxies at $z \sim 2$} \\
\noalign{\smallskip}
\hline
 \noalign{\smallskip}
9347 & $6.1^a$ & $9.5 \pm 3.0 $& $0.04 \pm 0.02 $ &$91.9 \pm 9.3$& $1.5 \pm 0.2 $  & [\dag, 15]\\ %
 \noalign{\smallskip}
6515 &$4.4^a$ &$12.3 \pm 3.3 $ & $0.10 \pm 0.05 $ & $65.8 \pm 5.6$ & $2.1 \pm 0.9 $  & [\dag, 15] \\ %
 \noalign{\smallskip}
10076 & $4.7^a$ &$24.0 \pm 7.6 $ & $0.2 \pm 0.1 $ & $121.1 \pm 7.2$ & $3.5 \pm 0.2 $  & [\dag, 15]\\ %
 \noalign{\smallskip}
9834 & $1.9^a$ & $12.9 \pm 1.9 $& $0.5 \pm 0.2 $ & $128.2 \pm 5.8$& $21.7 \pm 1.3$  & [\dag, 15] \\ %
 \noalign{\smallskip}
9681 & $3.8^a$&$18.1 \pm 7.9 $ & $0.2 \pm 0.1 $ & $95.4 \pm 8.2$  &$4.1 \pm 0.4 $  & [\dag, 15]\\ %
 \noalign{\smallskip}
8490 &$3.2^a$ &$7.1 \pm 3.0 $ & $0.11 \pm 0.06 $ & $53.0 \pm 6.1$ & $3.3 \pm 0.4 $ & [\dag, 15]\\ %
 \noalign{\smallskip}
\hline
 \noalign{\smallskip}
 \multicolumn{7}{c}{Literature sample of sources at $6< z < 7.5$} \\
 \noalign{\smallskip}
 \hline
 \noalign{\smallskip}
COS-3018  & $2.6 \pm 0.5$ & $4.7 \pm 0.5$&$0.11 \pm 0.04 $ & $18.9 \pm 1.5^c$  &$1.8 \pm 0.3$ &  [1, 5]  \\
 \noalign{\smallskip}
A383-5.2 &  $1.9^a$ &$0.09 \pm 0.03$ & $0.004 \pm 0.002 $& $2.2 \pm 0.6^c$ & $0.4 \pm 0.2$   &  [1, 11, 6] \\ %
 \noalign{\smallskip}
 z7\_GND\_42912 & $1.4 \pm 0.2^a$ &<3.6  & <0.3  & $14.7^c$ &$4.5 \pm 0.9$ & [\dag, 12, 16]  \\ %
 \noalign{\smallskip}
Himiko & $(3.9\pm1.1)\times(1.7\pm1.1)$ &$1.2 \pm 0.2 $ & $0.03 \pm 0.02 $  & $20.4 \pm 1.5^c$ &$4.0 \pm 1.6$ &  [4] \\ %
 \noalign{\smallskip}
CR7 &  ${3.8}^b $ & $2.0 \pm 0.4$& $0.02 \pm 0.01 $   & $28.4 \pm 1.3$  & $5.6 \pm 2.2$   & [1, 13] \\ %
 \noalign{\smallskip}
CR7a & $3.0$& $0.9 \pm 0.2 $& $0.02 \pm 0.01 $  & $17.6 \pm 0.6$ & $3.5 \pm 1.4$  & [1, 13]\\ %
 \noalign{\smallskip}
CR7b & <2.2& $0.3 \pm 0.1$ & $0.009 \pm 0.002 $  & $3.2_{-0.6}^{+1.3}$  & $0.6_{-0.3}^{+0.4}$  & [1, 13] \\  %
 \noalign{\smallskip}
CR7c &  $3.8$ & $0.3 \pm 0.1$ & $0.003 \pm 0.002 $   &$4.4 \pm 0.6$  & $0.9 \pm 0.4$    & [1, 13]\\ %
 \noalign{\smallskip}
COS-13679 & 1.4 &$0.7 \pm 0.2 $ &  $0.06 \pm 0.03 $ & $13.9^c$ & $3.5 \pm 1.5$ & [1, 14]\\ %
 \noalign{\smallskip}
COS-2987 & $3.1 \pm 1.0 $ & $3.6 \pm 0.5 $& $0.06 \pm 0.04 $  &$22.7 \pm 2.0^c$ &  $1.5 \pm 1.0$   & [5] \\ %
 \noalign{\smallskip}
VR7 & $(3.9\pm0.4) \times (1.9\pm0.2)$ &$4.8 \pm 0.4 $ & $0.10 \pm 0.02 $ &$29.7 \pm 1.1$ &$1.9 \pm 0.1$ & [6] \\ %
 \noalign{\smallskip}
\hline
\end{tabular}
\tablefoot{SFR reported in the literature are converted to the \cite{2012ARA&A..50..531K} scaling relations, when necessary. We assume 20\% uncertainties on the $r_{\rm{[\ion{C}{II}]}}$ and SFR when they are not provided. $^a$ [\ion{C}{II}] radius estimated as two times the UV radius. $^b$ [\ion{C}{II}] radius of the largest component of a multi-component system, $^c$ Total SFR is equal to the UV SFR.  

}
\tablebib{ [\ion{C}{II}] radius, luminosity, and SFR references: [\dag] This work, [1] \cite{2018MNRAS.478.1170C}, [2] \cite{2017ApJ...836L...2B}, [3] \cite{2010A&A...513A..20V}, [4] \cite{2018ApJ...854L...7C}, [5] \cite{2018Natur.553..178S}, [6] \cite{2019ApJ...881..124M}, [7]  \cite{2017MNRAS.466.3612B}, [8] \cite{2015Natur.522..455C}, [9]  \cite{2016ApJ...817...11H}, [10] \cite{2020A&A...643A...2B}, [11] \cite{2016MNRAS.462L...6K}, [12] \cite{2015A&A...574A..19S}, [13] \cite{2017ApJ...851..145M}, [14] \cite{2016ApJ...829L..11P}, [15]  \cite{2018MNRAS.481.1976Z}, [16]  \cite{2020MNRAS.494.1071G}.
}
\end{table*}

\subsection{ISM parameters: $\kappa_s$, $Z$, and $n$} \label{constraining}

In this section we exploit the \ion{C}{III}] and [\ion{C}{II}] emission to constrain the ISM properties of the selected sample. By adopting the model by \cite{2020MNRAS.495L..22V}, which depends on $\Sigma_{\ion{C}{III}]}$, $\Sigma_{[\ion{C}{II}]}$ and $\Sigma_{\rm SFR}$, we can determine the total gas density $n$, gas-phase metallicity $Z$, and the deviation from the KS relation, quantified by the burstiness parameter $\kappa_s$.

We initially applied the \cite{2020MNRAS.495L..22V} model on those galaxies that are detected both in \ion{C}{III}] and [\ion{C}{II}] emission. The results are reported in Table \ref{params}.
For the other sources with upper limits on either the \ion{C}{III}] or [\ion{C}{II}] emission, we constrained the ISM parameters, assuming that the upper limits are measured values with uncertainties equal to $25\%$ of the upper limit estimates. We report the derived values for these sources in Table \ref{params2} (Appendix \ref{app_params}). The detected high-$z$ galaxies have a gas-phase metallicity in the range  $0.2 \ Z_{\odot} \lesssim Z \lesssim 0.6 \ Z_{\odot}$ and a total gas density in the range $2.4 \lesssim \log{(n[\rm{cm^{-3}}])}\lesssim 3.4$. On average, all galaxies deviate from the KS relation by about one order of magnitude ($1.1\lesssim \log{(\kappa_s)} \lesssim 1.4$). This implies that these early galaxies are mostly in a starburst phase, with a higher star formation rate with respect to their gas content. Furthermore, the available data at $5<z<7$ suggests that the KS relation is not universal and it might be shifted upwards at higher redshift (\citealt{2021MNRAS.505.5543V}; \citealt{2022MNRAS.513.5621P}; see \citealt{2013ARA&A..51..105C} and \citealt{2015arXiv151103457K} for reviews).

\begin{table}[h]
 \caption[]{\label{params}Burstiness parameter $\kappa_s$, gas metallicity $Z$, and gas density $n$ for the three $z>5$ galaxies detected in both \ion{C}{III}] and [\ion{C}{II}].}
 \centering
\begin{tabular}{lcccccccc}
 \hline \hline
  \noalign{\smallskip}
  ID & $\log{(\kappa_s)}$ & $Z$ & $\log{(n)}$ \\
    \noalign{\smallskip}
  & & $(Z_{\odot})$ & $(\rm{cm^{-3}})$ \\
  \noalign{\smallskip}
 \hline
 \noalign{\smallskip}
HZ10 & $1.43_{-0.53}^{+0.39} $& $0.60_{-0.52}^{+0.32}$ & $3.35_{-0.15}^{+0.10}$\\
 \noalign{\smallskip}
COS-3018 & $1.09_{-0.13}^{+0.16}$& $0.34_{-0.43}^{+0.34}$ & $2.72_{-0.31}^{+0.55}$ \\
 \noalign{\smallskip}
A383-5.2  & $1.13_{-0.15}^{+0.29}$&  $0.24_{-0.40}^{+0.29}$ & $2.41_{-0.43}^{+0.62}$ \\ %
 \noalign{\smallskip}
\hline
\end{tabular}
\end{table}

We note that \cite{2019ApJ...882..168P} derived a lower $\kappa_s$ parameter  $\log{(\kappa_s)} \approx -1$ for HZ10, by observing the CO(2-1) line, which implies a low star formation efficiency for this source. The conflict between the two results can be explained by the fact that \cite{2019ApJ...882..168P} estimated the gas mass by adopting a large CO-to-$\rm{M_{gas}}$ conversion factor $\alpha_{\rm{CO}} = 4.5 \ M_{\odot} \ {(\rm{ K \ km \ s^{-1} pc^2}})^{-1}$, a value that is close to the Galactic conversion factor $\alpha_{\rm{CO}} = 4.36 \ M_{\odot} \ {(\rm{ K \ km \ s^{-1} pc^2}})^{-1}$ (\citealt{2013ARA&A..51..207B}). Although the Galactic conversion factor is a derived value for Milky Way and normal, star-forming galaxies in the local Universe, it may not be applicable for more extreme environments of starburst galaxies at high-$z$ (see \citealt{2013ARA&A..51..105C} for a review). The conversion factor depends on the physical conditions of the gas in the ISM (temperature, surface density, dynamics, and metallicity), as well as the star formation and associated feedback (\citealt{2011MNRAS.418..664N, 2012MNRAS.421.3127N}; \citealt{2012ApJ...746...69G}; \citealt{Feldmann_2012}; \citealt{2019A&A...621A.104R}; see, e.g., \citealt{2013ARA&A..51..207B} for a review). It is typically in the range between 0.8 and 4.36  $M_{\odot} \ {(\rm{ K \ km \ s^{-1} pc^2}})^{-1}$  (see, e.g., \citealt{2013ARA&A..51..207B}; \citealt{2013ARA&A..51..105C}, and \citealt{2018A&ARv..26....5C}  for a review). Low metallicities ($Z = 0.6 \ Z_{\odot}$ for HZ10) will drive $\alpha_{\rm{CO}}$ towards values higher than the Galactic value (\citealt{2012MNRAS.421.3127N}; \citealt{2012ApJ...746...69G}; \citealt{2014MNRAS.444.1301P}), although $\alpha_{\rm{CO}}$ spans a broad range of values of $ \alpha_{\rm{CO}} \sim 0.4-11 $ $M_{\odot} \ {(\rm{ K \ km \ s^{-1} pc^2}})^{-1}$, due to large uncertainties (see, e.g., Fig. 9 of \citealt{2013ARA&A..51..207B}). On the other hand, high values of temperature, surface density, and velocity dispersion in a turbulent ISM of starbursts and merging systems will shift $\alpha_{\rm{CO}}$ towards lower values (\citealt{2011MNRAS.418..664N, 2012MNRAS.421.3127N}; \citealt{2018MNRAS.473..271V}). HZ10 has an extremely high value of the burstiness parameter $\log{(\kappa_s)} \sim 1.4$ and high total density of the [\ion{C}{II}] emitting gas $\log{(n)} \sim 3.35 \ \rm{cm^{-3}}$, and it is also a multi-component system (\citealt{Jones_2017}, \citealt{2018MNRAS.478.1170C}). Thus, for this source we assumed $\alpha_{\rm{CO}} = 0.8 \ M_{\odot} \ {(\rm{ K \ km \ s^{-1} pc^2}})^{-1}$, usually adopted for starburst galaxies (e.g., \citealt{1998ApJ...507..615D}; \citealt{2013ARA&A..51..207B}). We obtained $\log{(\kappa_s)} = 0.53 \pm 0.34$, which is within the $2\sigma$ uncertainties of the $\log{(\kappa_s)} = 1.43_{-0.53}^{+0.38}$, estimated exploiting the \ion{C}{III}] emission and the \cite{2020MNRAS.495L..22V} model.

Finally, we also note that the best-fit results estimated for COS-3018 are different from those reported by  \cite{2020MNRAS.495L..22V}, even though both works are based on the same model and the \ion{C}{III}] and [\ion{C}{II}] emission. \cite{2020MNRAS.495L..22V} found that COS-3018 is a moderate starburst with $\log{(\kappa_s)} \approx 0.5$, with a mean gas density $\log{(n[\rm{cm^{-3}}])} \approx 2.7$ and a gas metallicity of $Z \approx 0.4 \ Z_{\odot}$. The derived ISM properties are consistent with our values, apart from the burstiness parameter, which is $\log{(\kappa_s)} \approx 1.1$. The high value of the $\kappa_s$ parameter derived in our work is mainly driven by the value of the intrinsic \ion{C}{III}]  flux which is by a factor of $\sim 6.5$ higher, namely, $F_{\ion{C}{III}]} \approx 8.5 \times 10^{-18} \rm{erg \ s^{-1} \ cm^{-2}}$. On the other hand, \cite{2020MNRAS.495L..22V} used a lower value of the observed \ion{C}{III}] flux of $F_{\ion{C}{III}]} \approx 1.3 \times 10^{-18} \rm{erg \ s^{-1} \ cm^{-2}}$ (\citealt{2017ApJ...851...40L}), without considering the correction for the dust attenuation. 


\section{Discussion} \label{Discussion}

In this section, we analyze the distribution of high-$z$ galaxies in the $\Sigma_{[\ion{C}{II}]}-\Sigma_{\rm{SFR}}$ space and  compare it with the local relation from \cite{2014A&A...568A..62D} and recent theoretical models.
Next, we investigate the possible correlation between the ISM properties: $\kappa_s$, $Z$, and $n$, as well as the deviation from the local relation, namely, the $\Delta$ parameter. 

\subsection{$\Sigma_{[\ion{C}{II}]}$-$\Sigma_{\rm{SFR}}$ relation} \label{deviation}

\cite{2014A&A...568A..62D} found a tight relation between the [\ion{C}{II}] emission and SFR for a sample of spatially resolved,  low-metallicity dwarf galaxies in the local Universe, which can be expressed as:

\begin{equation} \label{delooze}
   \rm{\log{(\Sigma_{SFR})}} = -6.99 + 0.93\ \rm{\log {(\Sigma_{[\ion{C}{II}]})}}.
\end{equation}%

\begin{figure}
\centering
\includegraphics[width=\hsize]{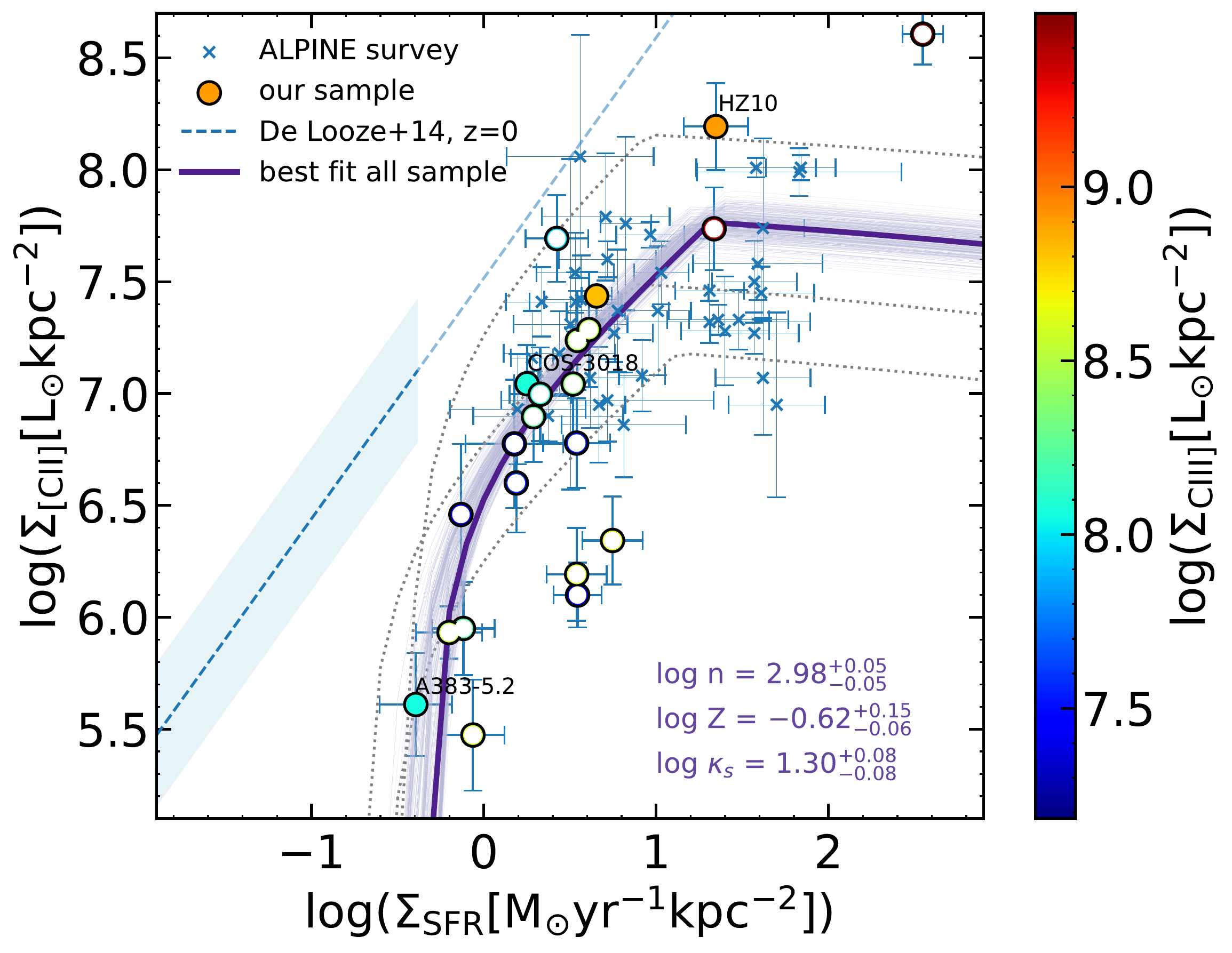}
 \caption{[\ion{C}{II}] surface brightness versus SFR per unit area, color-coded by the \ion{C}{III}] surface brightness, of this sample of galaxies. Filled and empty circles indicate sources with the \ion{C}{III}] line detection and upper limits, respectively. Galaxies detected both in [\ion{C}{II}] and \ion{C}{III}] emission are annotated. Blue $\times$ symbols indicate a subsample of galaxies from the ALPINE survey at $4 < z < 6$ (\citealt{2020A&A...643A...1L}).  $\Sigma_{[\ion{C}{II}]}-\Sigma_{\rm{SFR}}$ relation for local galaxies by \cite{2014A&A...568A..62D} is represented as a blue dashed line with a $1\sigma$ dispersion of 0.32 dex. This relation is extrapolated to cover the distribution of high-$z$ sources in $\Sigma_{[\ion{C}{II}]}-\Sigma_{\rm{SFR}}$ space, which is shown as a light-blue dashed line. $\Sigma_{[\ion{C}{II}]}-\Sigma_{\rm{SFR}}$ relation predictions from models of \cite{2019MNRAS.489....1F} with three different combinations of fixed $\kappa_s$, $Z$, and $n$ parameters (Table \ref{params}) are represented as dotted gray lines. Finally, the best-fit model ($Z = 0.24 \ Z_{\odot}$, $\log{(n[\rm{cm^{-3}}])} =  2.98$, and $\log{(\kappa_s)} \approx 1.30$) on all the galaxies (excluding Z7\_GND\_42912) and 100 random samples from the MCMC chain are shown as thick and thin purple solid lines, respectively.          }
     \label{De-Looze}
\end{figure}

We investigate whether this relation is still valid in the high-$z$ Universe. In Fig. \ref{De-Looze}, we show the [\ion{C}{II}] surface brightness and SFR per unit area of our entire sample of high-$z$ ($5.5 < z < 7.5$) and intermediate-$z$ ($z \sim 2$) galaxies. To increase our sample and cover a broader range of redshift as well as the $\Sigma_{[\ion{C}{II}]}-\Sigma_{\rm{SFR}}$ parameter space, we also include a subsample of 38 galaxies from the ALPINE survey at $4 < z < 6$, for which [\ion{C}{II}] measurements and radii are well constrained and available in the literature (\citealt{2020A&A...643A...1L}; \citealt{2020ApJ...900....1F}; \citealt{2020A&A...643A...2B}) and we derived their total SFR from the UV and IR luminosities (if the galaxies are detected in the IR). 

All of the sources lie below the $\Sigma_{[\ion{C}{II}]}-\Sigma_{\rm{SFR}}$ local relation by \citealt{2014A&A...568A..62D} (Fig. \ref{De-Looze}, blue dashed line). This is consistent with the results of \cite{2018MNRAS.478.1170C}  in a sample of $z >5$ star-forming galaxies. We note that part of the \citealt{2014A&A...568A..62D} local relation in Fig. \ref{De-Looze} is an extrapolation of the actual relation found by \citealt{2014A&A...568A..62D},  since their sample of spatially resolved dwarf galaxies covers a lower range of values in the $\Sigma_{[\ion{C}{II}]}-\Sigma_{\rm{SFR}}$ space, with $4.5 \lesssim \log ({(\Sigma_{[\ion{C}{II}]})}[L_{\odot} \ kpc^{-2}]) \lesssim 7.0$ and $-3 \lesssim \rm{\log{(\Sigma_{SFR}[M_{\odot} \ yr^{-1}])}} \lesssim 0$, compared to most high-$z$ sources used in this work.  

In Fig. \ref{De-Looze}, we report three $\Sigma_{[\ion{C}{II}]}-\Sigma_{\rm{SFR}}$ models by \cite{2019MNRAS.489....1F} that fit the three high-$z$ sources with both $\ion{C}{III}]$ and $[\ion{C}{II}]$ detections (dotted gray lines), for which the ISM properties are constrained in Sect. \ref{constraining} (Table \ref{params}). We can notice that the model with fixed ISM properties of the HZ10 source ($\log{(\kappa_s)} \approx 1.43$, $Z = 0.6 \ Z_{\odot}$, and $\log{(n[\rm{cm^{-3}}])} =  3.35$), passes through HZ10 and A383-5.2. However, this model overestimates the $\kappa_s$ parameter in A383-5.2, because $\kappa_s$ is mainly driven by very high \ion{C}{III}] emission, which is not observed in A383-5.2. Therefore, the model with fixed ISM properties of the A383-5.2 source ($\log{(\kappa_s)} \approx 1.13$, $Z = 0.24 \  Z_{\odot}$, and $\log{(n[\rm{cm^{-3}}])} =  2.41$) is clearly a better fit for the A383-5.2 galaxy.

Next, we show the best-fit model of our full sample (excluding Z7\_GND\_42912 as it is not detected in [\ion{C}{II}]) and the ALPINE subsample (Fig. \ref{De-Looze}, solid purple line).  The best-fit model for these high-$z$ sources points to reasonable average values of subsolar metallicity $Z = 0.24 \ Z_{\odot}$, high density $\log{(n[\rm{cm^{-3}}])} =  2.98$, and a very high burstiness parameter $\log{(\kappa_s)} \approx 1.3$ with respect to the local dwarf sample from \cite{2014A&A...568A..62D}. The high mean value of the $\kappa_s$ parameter is another strong piece of evidence in favor of the KS relation being shifted upwards at higher-$z$ (\citealt{2021MNRAS.505.5543V}; \citealt{2022MNRAS.513.5621P}). We will discuss this aspect in more detail in an upcoming work. 

Overall, these models show that the [\ion{C}{II}] faintness of high-z star-forming galaxies might be driven by a combination of high burstiness parameter, low metallicity, and total gas density. Clearly, it would be possible to fit a single source with multiple models since there is a degeneracy between the ISM parameters: $Z$, $n$, and $\kappa_s$. Therefore, besides the [\ion{C}{II}] line data, observations of an additional line tracing the ISM is necessary to resolve the degeneracy between these three parameters. 

\begin{figure}
\centering
\includegraphics[width=\hsize]{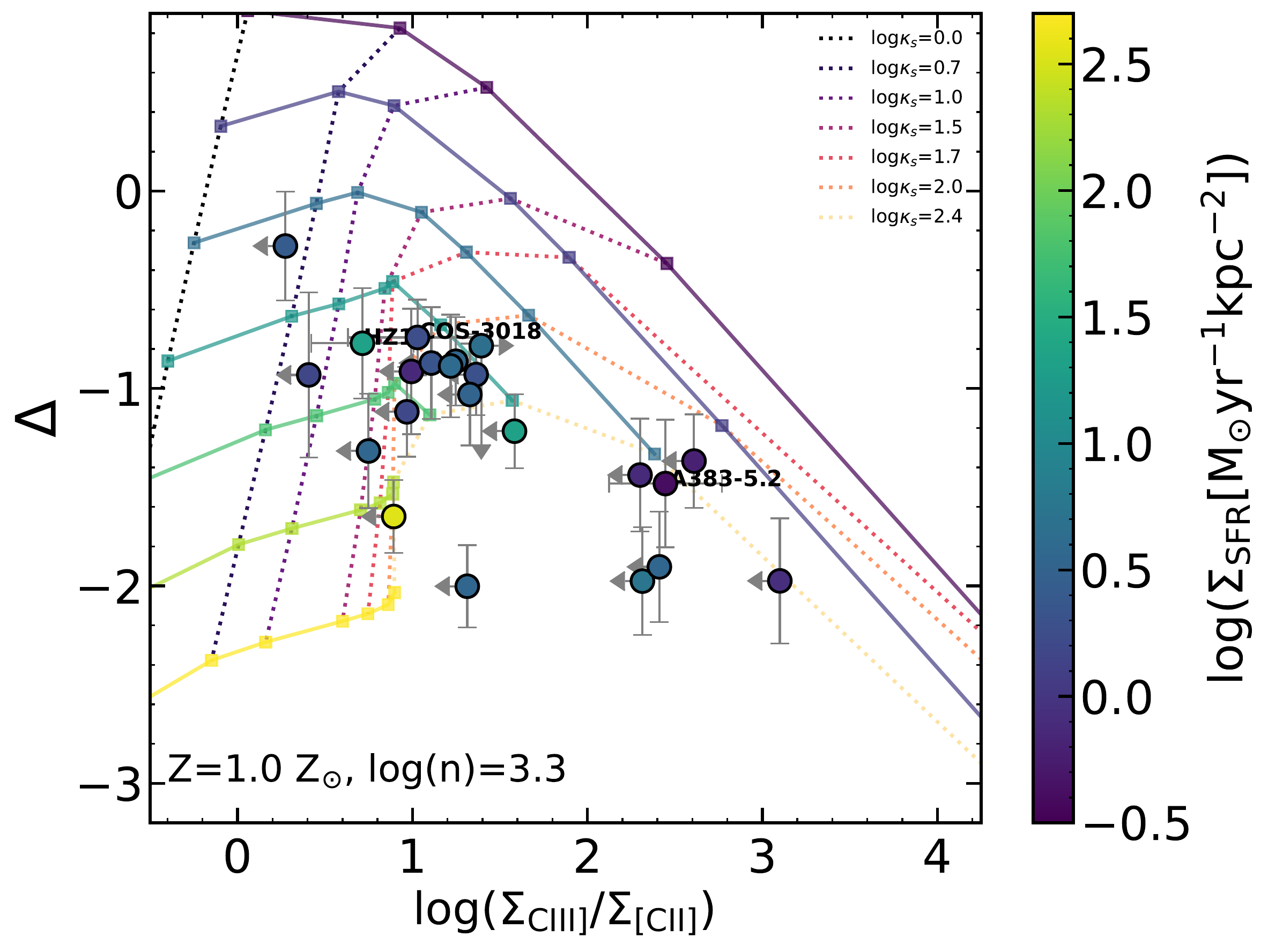}
\includegraphics[width=\hsize]{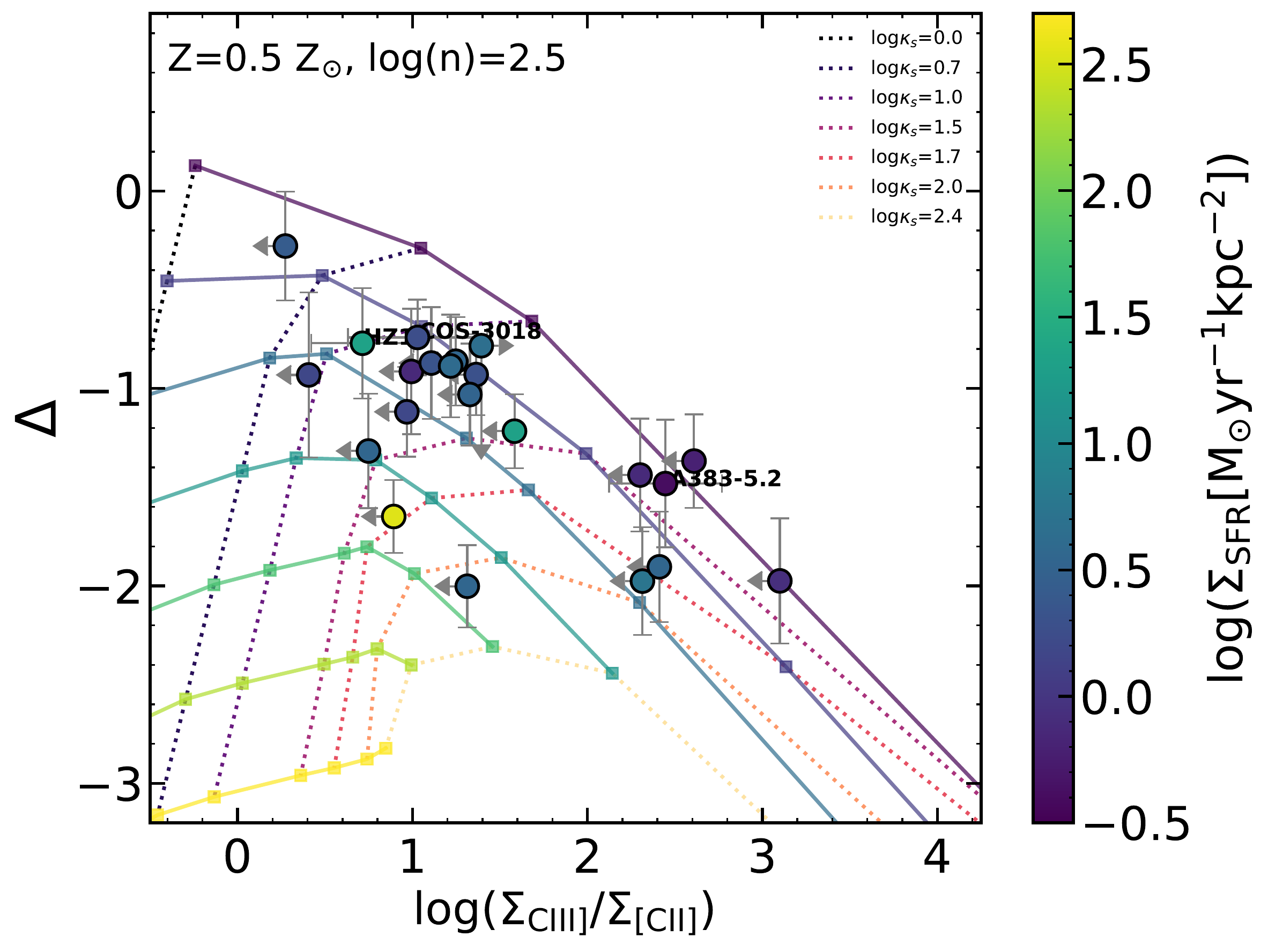}
 \caption{Deviation from the local $\Sigma_{[\ion{C}{II}]}-\Sigma_{\rm{SFR}}$ relation $\Delta$ vs. the \ion{C}{III}] over [\ion{C}{II}] surface brightness ratio, color-coded by the SFR per unit area $\Sigma_{\rm{SFR}}$. Galaxies detected both in [\ion{C}{II}] and \ion{C}{III}] emission are annotated. We plot a grid of model predictions with fixed metallicity and gas density, with the burstiness parameter varying in a range of $0 \le \log{(\kappa_s)} \le 2.4$  i.e., $\kappa_s = 1, 5, 10, 30, 50, 100, 250$ (dotted lines), and $\log{(\Sigma_{\rm{SFR}}[M_{\odot} \ yr^{-1}])}$ in  $-0.5 \le \rm{\log{(\Sigma_{SFR}[M_{\odot} \ yr^{-1}])}} \le 2.5$ (solid lines). The top panel shows an example of a model prediction for high-$z$ galaxies with ISM properties similar to HZ10, with solar metallicity $Z = Z_{\odot}$ and gas density close to critical density of $\log{(n[\rm{cm^{-3}}])} = 3.3$. Bottom panel shows predictions for typical high-$z$ galaxies with subsolar metallicity $Z = 0.5\ Z_{\odot}$ and $\log{(n[\rm{cm^{-3}}])} = 2.5$.               }
     \label{delta_ratio}
\end{figure}

In Fig. \ref{delta_ratio}, we show the deviation from the local $\Sigma_{[\ion{C}{II}]}-\Sigma_{\rm{SFR}}$ relation $\Delta$ and the \ion{C}{III}] over [\ion{C}{II}] surface brightness ratio, color-coded by the total SFR per unit area $\Sigma_{\rm{SFR}}$. The $\Delta$ parameter is generally decreasing, namely, $|\Delta|$ is increasing, with the increase of the $\Sigma_{\ion{C}{III}]}/\Sigma_{[\ion{C}{II}]}$ ratio and $\Sigma_{\rm{SFR}}$, as expected from the model of \cite{2020MNRAS.495L..22V}. We overplot the grid of model predictions with fixed $Z$ and $n$,  allowing for $\kappa_s$ and  $\Sigma_{\rm{SFR}}$ to vary in a broad range of values of $0 \le \log{(\kappa_s)} \le 2.4$ (Fig. \ref{delta_ratio}, dashed lines) and $-0.5 \le \rm{\log{(\Sigma_{SFR}[M_{\odot} \ yr^{-1}])}} \le 2.5$ (Fig. \ref{delta_ratio}, solid lines), respectively. In the top panel of Fig. \ref{delta_ratio}, we show a model example with a solar metallicity $Z = Z_{\odot}$ and total gas density of $\log{(n[\rm{cm^{-3}}])} = 3.3$, equivalent to the HZ10 example model shown in Fig. \ref{De-Looze} (dashed purple line). Moreover, in the bottom panel of Fig. \ref{delta_ratio}, we show an example with $Z = 0.5\ Z_{\odot}$ and $\log{(n[\rm{cm^{-3}}])} = 2.5$, typical values for high-$z$ galaxies (\citealt{2017MNRAS.471.4128P}). We can observe that the first model correctly reproduces the $\Sigma_{\rm{SFR}}$ parameters for the HZ10 source, whereas the second model reproduces correctly most of the other galaxies from our sample (Fig. \ref{delta_ratio}, top and bottom panel, respectively), implying that a subsolar metallicity is a more plausible average value for the large majority of high-$z$ star-forming galaxies.

 \subsection{Correlation between the deviation and ISM properties} \label{correlation}

We now assess whether the [\ion{C}{II}] deficit depends on the ISM properties. In particular, we investigate the relation between the $\Delta$ parameter and ISM properties, focusing mostly on the $\kappa_s$ and $n$, because $Z$ is not well constrained by the model. 

 Since, in our sample, we have only three galaxies for which  $\kappa_s$ and $n$ are well constrained, in this part of our analysis, we include a sample of galaxies in the EoR (at $6 < z < 9$) with $\kappa_s$, $Z$, and $n$ constrained by \cite{2021MNRAS.505.5543V}\footnote{GLAM (Galaxy Line Analyzer with MCMC) code is publicly released on GitHub (\url{https://lvallini.github.io/MCMC_galaxyline_analyzer/})}. This method is equivalent to the \cite{2020MNRAS.495L..22V} method used in this work, with the exception of a different line used as a ionized gas tracer, namely: the [\ion{O}{III}] line at $\lambda = 88 \ \rm{\mu m}$.

\begin{figure}
\centering
\includegraphics[width=\hsize]{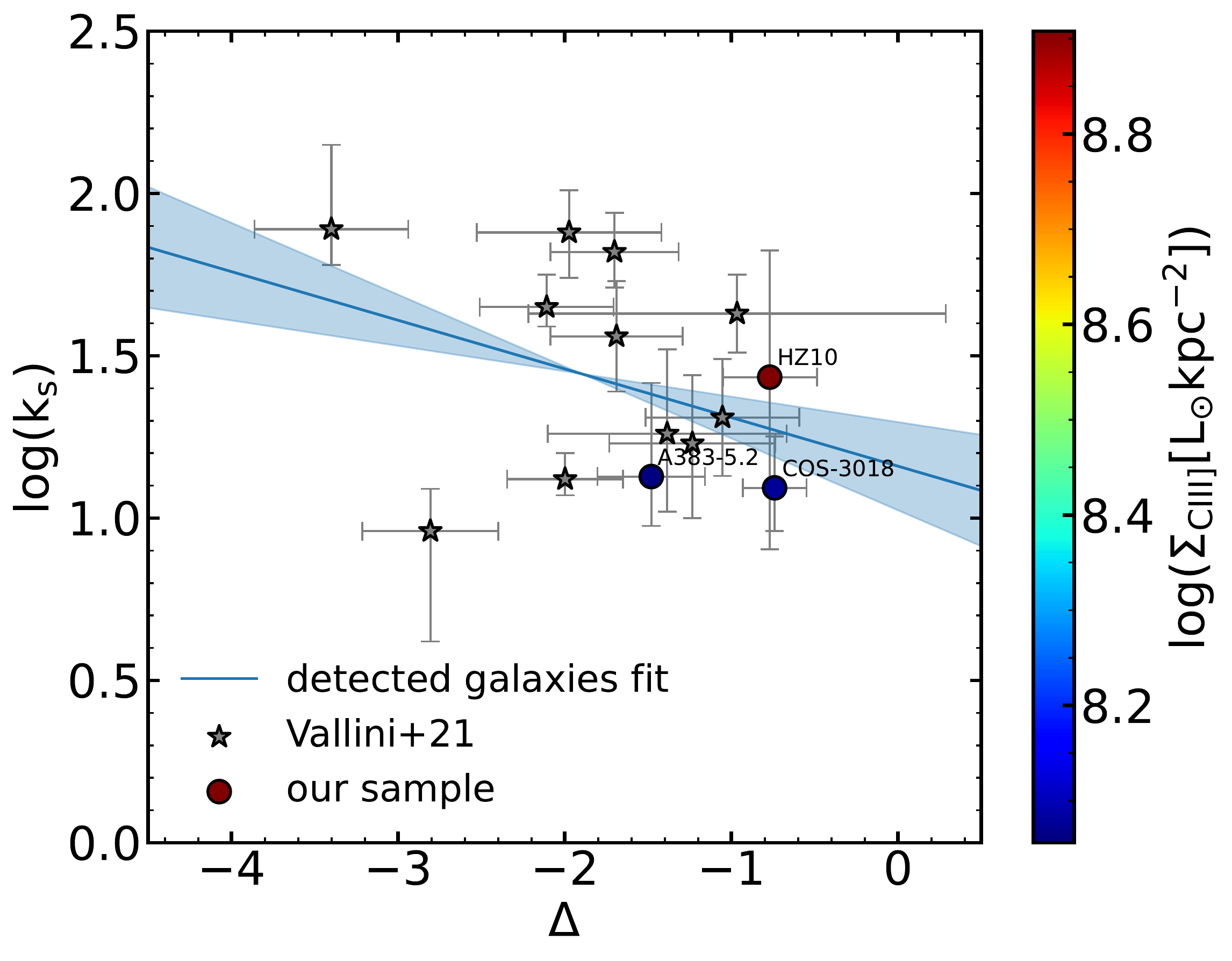}
 \caption{Burstiness parameter $\kappa_s$ vs. deviation from the local De-Looze relation $\Delta$, color-coded by the  \ion{C}{III}] surface brightness $\Sigma_{ \ion{C}{III}]}$. Galaxies detected both in [\ion{C}{II}] and \ion{C}{III}] emission are annotated and represented as filled circles. Galaxies at $6<z<9$ from \cite{2021MNRAS.505.5543V} are represented as gray stars. Linear fit on the detected sources is shown as a blue line with $1\sigma$ uncertainties as a shaded region.
          }
     \label{ks_delta}
\end{figure}

In Fig. \ref{ks_delta}, we report the burstiness parameter $\kappa_s$ and the deviation from the  $\Sigma_{[\ion{C}{II}]}-\Sigma_{\rm{SFR}}$ relation $\Delta$. Next, we explore the correlation between the $\kappa_s$ and $\Delta$ by using the  Spearmann rank correlation, resulting into a correlation coefficient of $\rho \approx -0.27$ and the $p$-value of $p=0.35$. Thus, current data seem to indicate that the [\ion{C}{II}] deficit is anti-correlated to the burstiness parameter, but it is not statistically significant. From the best-fit on the detected $z>5$ sources we see that $\kappa_s$ is increasing with the decrease of the $\Delta$ parameter, that is, as galaxies undergo a larger deviation from the local De-Looze relation,  $|\Delta|$ is increasing if they are in a more "bursty" phase (Fig. \ref{ks_delta}, blue line). This is expected, since intense radiation fields emitted by young massive stars in starburst galaxies with $\kappa_s \gg 1$ are strong enough to photo-evaporate molecular clouds (\citealt{2019MNRAS.487.3377D}) and the PDRs bound to them, from where the bulk of the [\ion{C}{II}] emission originates (\citealt{2015ApJ...813...36V, 2017MNRAS.467.1300V}). Consequently, strong UV radiation will alter the ionization state of the bulk of the gas, including carbon, which will emit more in \ion{C}{III}] or higher ionization states (e.g., \citealt{2019MNRAS.489....1F}; \citealt{2020MNRAS.495L..22V}). Therefore, these starbursts are expected to deviate from the local $\Sigma_{\rm{[\ion{C}{II}]}}-\Sigma_{\rm{SFR}}$ relation.

\begin{figure}
\centering
\includegraphics[width=\hsize]{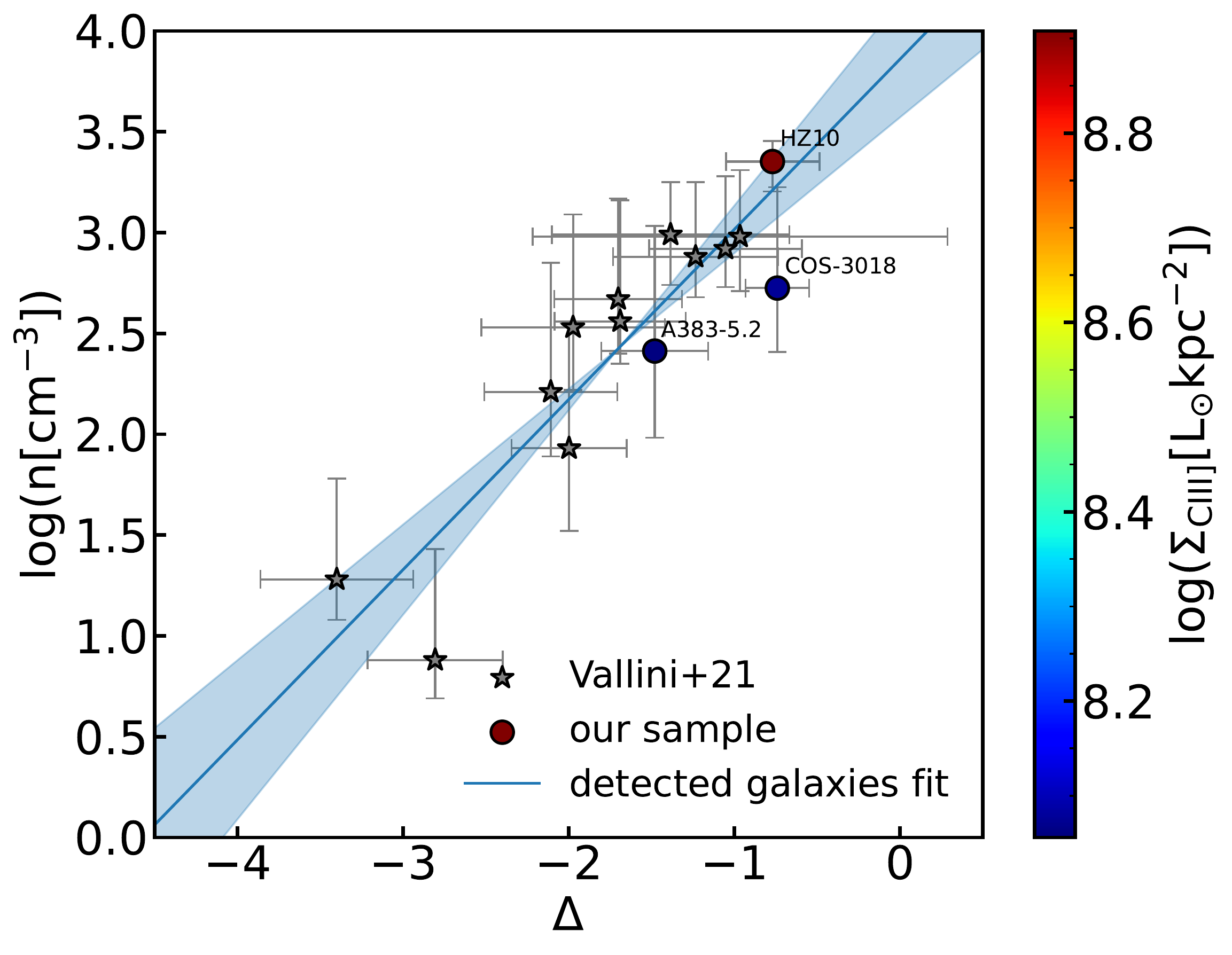}
 \caption{Density of the emitting gas $n$ vs. deviation from the local De-Looze relation $\Delta$, color-coded by the \ion{C}{III}] surface brightness $\Sigma_{\ion{C}{III}]}$. Symbols are the same as in Fig. \ref{ks_delta}.
          }
     \label{n_delta}
\end{figure}

Figure \ref{n_delta} illustrates the total gas density $n$ as a function of the  deviation from the local De-Looze relation $\Delta$ for high-$z$ galaxies. We derive the Spearmann correlation coefficient of $\rho \approx 0.86$ and the $p$-value of $p = 5.68 \times 10^{-5}$. Thus, the total gas density, $n,$ shows the strongest correlation with the $\Delta$ parameter out of any ISM property considered in this work. The best fit on the detection points shows that sources with low density, $n,$ tend to experience a larger deviation from the local De-Looze relation. This agrees with the interpretation of \cite{2019MNRAS.489....1F}, who found that extremely diffuse high-$z$ systems  with $n \approx 20 \ \rm{cm^{-3}}$ will have a low [\ion{C}{II}] surface brightness (e.g., CR7c with $\Sigma_{[\ion{C}{II}]} \sim 10^6 \ L_{\odot} \ \rm{kpc^{-2}}$). This is primarily a consequence of the fact that the  
[\ion{C}{II}] emitting region can be spatially extended than their star-forming region (\citealt{2018MNRAS.478.1170C}; \citealt{2019ApJ...887..107F}). The extended [\ion{C}{II}] emission may originate from the outflowing [\ion{C}{II}] emitting gas or the [\ion{C}{II}] circumgalactic medium, whereas the star-forming region, traced by the rest-frame UV emission, may be heavily dust-obscured (\citealt{2018MNRAS.478.1170C}; \citealt{2018MNRAS.473.1909G}; \citealt{2019ApJ...887..107F}).  Furthermore, low density systems at high-$z$ tend to experience more CMB suppression effects on the [\ion{C}{II}] line emission (\citealt{2019MNRAS.487.3007K}). Finally, in low-density systems, FUV photons may be able to penetrate deeper into the ISM and ionize more of its gas, which will lead to carbon emitting less in [\ion{C}{II}] and more in higher ionization states.

\section{Summary and conclusions} \label{Conclusions}

In this work, we present an analysis of the novel VLT/X-SHOOTER observations targeting the \ion{C}{III}] line emission in three high-$z$ galaxies: HZ1 at $z \sim 5.69$, HZ10 at $z \sim 5.66$, and RXJ 1347-1216 at $z \sim 6.77$. We have included the archival X-SHOOTER data of two other sources: GDS3073 at $z\sim 5.56$ and UVISTA-Z-002 at $z \sim 6.63$. We have also included the VLT/MUSE archival data of six galaxies at $z \sim 2$. We have extended our sample of galaxies with 11 star-forming systems at $6.0 < z< 7.5$ with \ion{C}{III}] and [\ion{C}{II}] detection or upper limits reported in the literature. We have combined the data on the \ion{C}{III}] surface brightness $\Sigma_{\ion{C}{III}]}$, [\ion{C}{II}] surface brightness  $\Sigma_{[\ion{C}{II}]}$, and SFR per unit area $\Sigma_{\rm{SFR}}$ with the model of \cite{2020MNRAS.495L..22V}, to characterize the ISM properties of high-$z$ galaxies: the burstiness parameter, $\kappa_s$, metallicity, $Z$, and total gas density, $n$. Finally, we have investigated the correlation of the different ISM properties with the deviation from the local $\Sigma_{[\ion{C}{II}]}-\Sigma_{\rm{SFR}}$ relation found by \citealt{2014A&A...568A..62D}. Our main results are: 
 
   \begin{enumerate}
      \item The X-SHOOTER observations reveal a line emission associated with the blended \ion{C}{III}] $\lambda \lambda1907, 1909$ doublet line in the HZ10 source. We extracted the 1D spectrum and measured an integrated flux of the \ion{C}{III}] $\lambda \lambda1907, 1909$ line of $\rm{8.8 \pm 2.1 \times 10^{-17} \ erg \ s^{-1} \ cm^{-2}}$. We derived the $\ion{C}{III}]\ \lambda1909$ integrated flux of $\rm{3.7 \pm 1.1 \times 10^{-17} \ erg \ s^{-1} \ cm^{-2}}$,  assuming the $[\ion{C}{III}]  \ \lambda1907/\ion{C}{III}] \ \lambda1909$ line ratio of $1.4 \pm 0.2$. For the other two targets: HZ1, RXJ 1347-1216, and other galaxies drawn from the archive, we placed $3\sigma$ upper limits on the $\ion{C}{III}]$ line flux (Table \ref{CIII-prop}).
      \item  We correct the observed \ion{C}{III}] line flux for the dust attenuation, using the Calzetti attenuation law (\citealt{Calzetti_2000}), for HZ10 and a handful of other $z>6$ galaxies: COS-3018, A383-5.2 and Z7\_GND\_42912, with \ion{C}{III}] emission reported in the literature. For HZ10, we derived the intrinsic \ion{C}{III}] line flux of $\rm{6.5 \pm 2.0 \times 10^{-17} \ erg \ s^{-1} \ cm^{-2}}$.
      \item We  derived an integrated flux of $35.5 \pm 7.7 \ \rm{mJy \ beam^{-1} \ km \ s^{-1}}$ of the [\ion{C}{II}] line emission for the UVISTA-Z-002 galaxy. For the other galaxies from our sample, we included the [\ion{C}{II}] line luminosity from the literature. For our analysis, we also included the total SFR data, either from the literature or derived from the UV and IR luminosities (Table \ref{CII-SFR}).   
       \item We characterized the burstiness parameter, $\kappa_s$, total gas density, $n$, and metallicity, $Z$,  with the use of the \ion{C}{III}] and [\ion{C}{II}] emission and  \cite{2020MNRAS.495L..22V} model.  High-$z$ star-forming galaxies predominantly show subsolar metallicities. The majority of all the high-$z$ sources are in the starburst phase with $\log{(\kappa_s)} \gtrsim 1$, implying that the $\Sigma_{\rm{SFR}}-\Sigma_{\rm{gas}}$ relation is shifted upwards at early epochs. 
       \item  We constrain the burstiness parameter $\log{(\kappa_s)} = 1.43_{-0.53}^{+0.38}$ for HZ10. This is not consistent with $\log{(\kappa_s)} \approx -1$, found by \cite{2019ApJ...882..168P}.  Using CO measurements of \cite{2019ApJ...882..168P} and assuming a lower conversion factor $\alpha_{\rm{CO}} = 0.8 \ M_{\odot} \ {(\rm{ K \ km \ s^{-1} pc^2}})^{-1}$, we obtain $\log{(\kappa_s)} = 0.53 \pm 0.34$. This value is approximately within the $2\sigma$ uncertainties of the $\log{(\kappa_s)}$ estimated using the \ion{C}{III}] emission and the \cite{2020MNRAS.495L..22V} model.
      \item We confirm that the high-$z$ galaxies generally lie below the $\Sigma_{\rm{[\ion{C}{II}]}}-\Sigma_{\rm{SFR}}$ local relation by \cite{2014A&A...568A..62D}, in agreement with the literature. The $\Sigma_{\rm{[\ion{C}{II}]}}-\Sigma_{\rm{SFR}}$ relation for high-$z$ galaxies  deviates from the simple power law and saturates at $\rm{\log(\Sigma_{[\ion{C}{II}]}[L_{\odot} \ kpc^{-2}])} \approx 7-8$, in agreement with the models from \cite{2019MNRAS.489....1F} and simulations from \cite{2019MNRAS.487.1689P}.
      \item Current observations show a weak anti-correlation between the deviation from the $\Sigma_{\rm{[\ion{C}{II}]}}-\Sigma_{\rm{SFR}}$ relation and the $\kappa_s$ parameter. Galaxies that are more "bursty" undergo a larger deviation from the local De Looze relation, which is in agreement with theoretical models.
       \item The total gas density, $n,$ shows the strongest correlation with the deviation from the local  $\Sigma_{\rm{[\ion{C}{II}]}}-\Sigma_{\rm{SFR}}$ relation. Sources with low total gas density, $n,$ tend to undergo a larger deviation from the local De Looze relation, in agreement with the interpretation of \cite{2019MNRAS.489....1F} for extremely diffuse high-$z$ galaxies.
     
   \end{enumerate}

In the near future, the synergy between ALMA and {\it JWST} will provide observations of [\ion{C}{II}] line  and nebular emission lines, including the \ion{C}{III}] line, of a large sample of the first galaxies in the EoR. These data will allow us to characterize the ISM properties of these galaxies and shed light on the origin of the deviation from the local $\Sigma_{\rm{[\ion{C}{II}]}}-\Sigma_{\rm{SFR}}$ relation at these early cosmic epochs.

\begin{acknowledgements}
       VM would like to thank A. Zanella, E. Ntormousi, F. Vito, L. Sommovigo, and A. Roy for their useful comments. VM, SC, LV, AF, and AP acknowledge support from the ERC Advanced Grant INTERSTELLAR H2020/740120. Any dissemination of results must indicate that it reflects only the author’s view and that the Commission is not responsible for any use that may be made of the information it contains. Partial support from the Carl Friedrich von Siemens-Forschungspreis der Alexander von Humboldt-Stiftung Research Award (AF) is kindly acknowledged. We gratefully acknowledge computational resources of the Center for High Performance Computing (CHPC) at SNS. R.M. acknowledges support by the Science and Technology Facilities Council (STFC) and ERC Advanced Grant 695671 "QUENCH". RM also acknowledges funding from a research professorship from the Royal Society. The work of this paper is based on observations with the X-SHOOTER and MUSE instruments of the ESO VLT. This work is based on data products from observations made with ESO Telescopes at La Silla Paranal Observatory under ESO programmes ID  0103.A-0692(A), 384.A-0886(A), 089.A-0679(A), 097.A-0082(A), 094.A-0289(B), 095.A-0010(A), 096.A-0045(A), 099.A-0060(A), 098.A-0017(A), 096.A-0090(A), 0100.D-0807(C), 095.A-0240(A). This paper makes use of the following ALMA data: ADS/JAO.ALMA\#2019.1.01634.L. ALMA is a partnership of ESO (representing its member states), NSF (USA) and NINS (Japan), together with NRC (Canada), MOST and ASIAA (Taiwan), and KASI (Republic of Korea), in cooperation with the Republic of Chile. The Joint ALMA Observatory is operated by ESO, AUI/NRAO, and NAOJ. This research has made use of "Aladin sky atlas" developed at CDS, Strasbourg Observatory, France (\citealt{2000A&AS..143...33B}; \citealt{2014ASPC..485..277B}).
\end{acknowledgements}

%
%
\bibliographystyle{aa} 
\bibliography{biblio} 

\begin{appendix} 

\section{NIR spectroscopy of galaxies}

\begin{figure}
\centering
\includegraphics[width=\hsize]{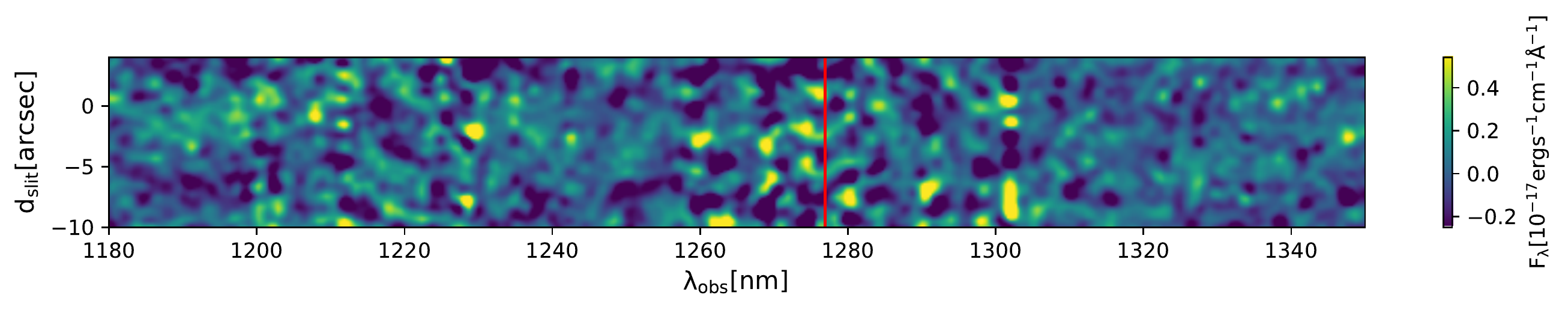}
\includegraphics[width=\hsize]{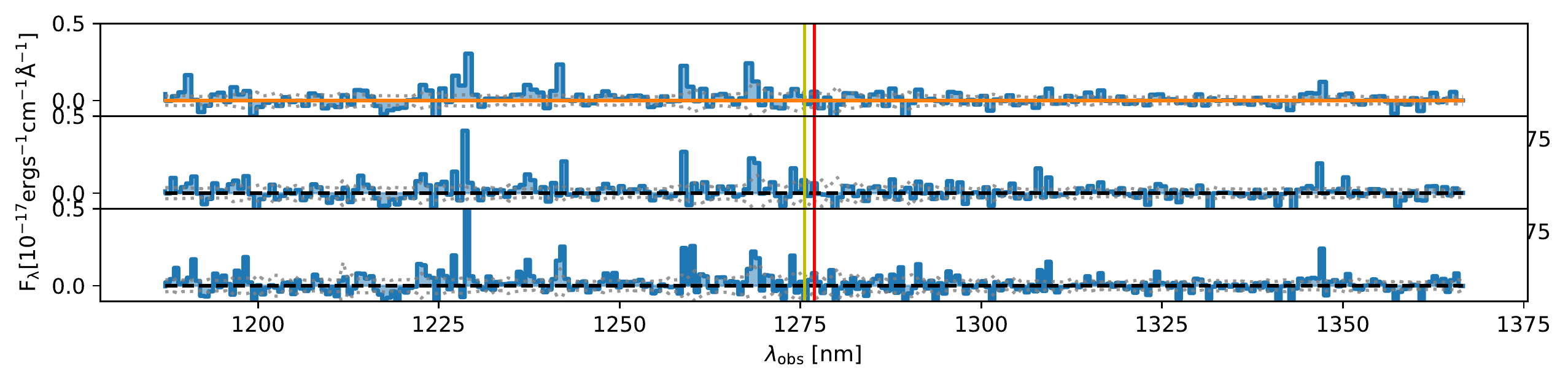}
 \caption{X-SHOOTER  NIR spectrum of the HZ1 galaxy. Top panel shows the combined, rebinned ($\sim 85 \ \rm{km \ s^{-1}}$), and smoothed ($\rm{0.7 \ nm \times 0.12 \arcsec}$)  2D spectrum. The bottom three panels illustrate the extracted 1D spectra rebinned by $\sim 210\ \rm{km \ s^{-1}}$,  $\sim 170\ \rm{km \ s^{-1}}$, and $\sim 140\ \rm{km \ s^{-1}}$, respectively. Expected positions of the \ion{[C}{III]} $\lambda 1907$ at  $\lambda = 1276.9$ nm and \ion{C}{III]} $\lambda 1909$ line at  $\lambda = 1275.6$ nm, based on the Ly$\alpha$ spectroscopic redshift, are shown as yellow and red lines, respectively. Shaded blue regions indicate the flux, dotted lines indicate the uncertainties, whereas the orange solid line is the best Gaussian fit.  
          }
     \label{2D-HZ1}
\end{figure}

\begin{figure}
\centering
\includegraphics[width=\hsize]{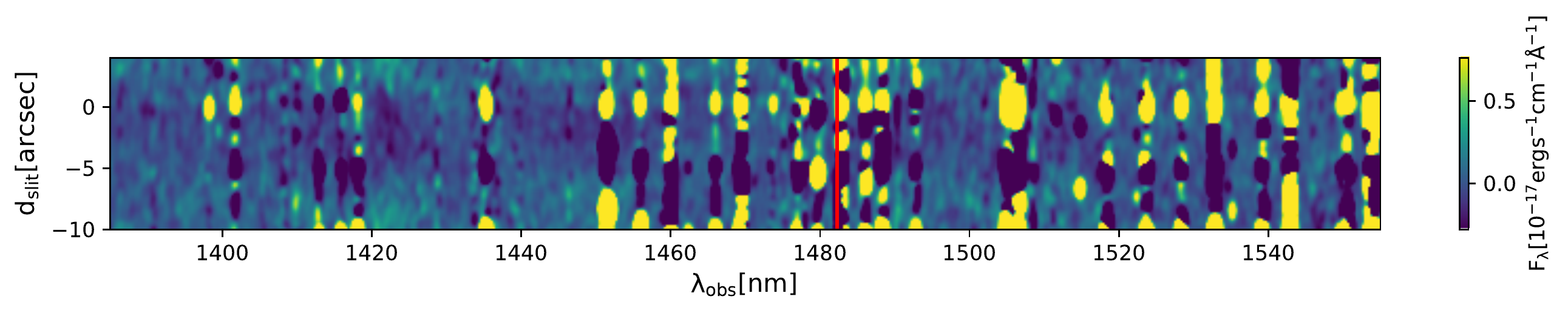}
\includegraphics[width=\hsize]{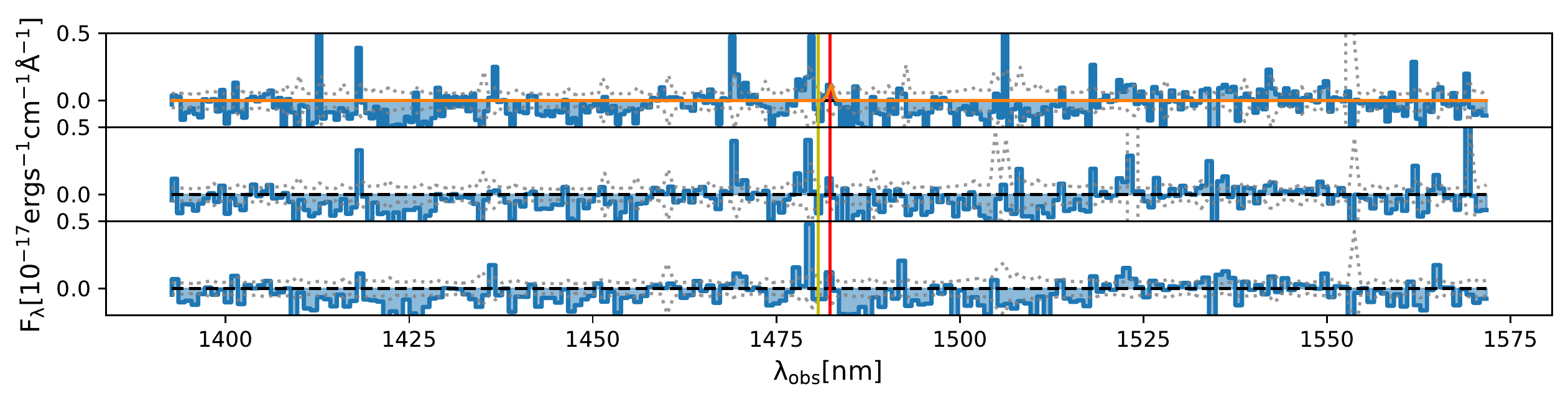}
 \caption{X-SHOOTER NIR spectra of the RXJ 1347-1216 galaxy. Top panel shows the combined, rebinned ($\sim 75 \ \rm{km \ s^{-1}}$), and smoothed ($\rm{0.7 \ nm \times 0.12 \arcsec}$)  2D spectrum. The bottom three panels illustrate the extracted 1D spectra rebinned by $\sim 180\ \rm{km \ s^{-1}}$,  $\sim 145\ \rm{km \ s^{-1}}$, and $\sim 120\ \rm{km \ s^{-1}}$, respectively. Expected positions of the \ion{[C}{III]} $\lambda 1907$ at  $\lambda = 1480.7$ nm and \ion{C}{III]} $\lambda 1909$ line at  $\lambda = 1482.3$ nm, based on the Ly$\alpha$ spectroscopic redshift, are shown as yellow and red lines, respectively. Labels are the same as in Fig. \ref{2D-HZ1}.
          }
     \label{2D-RXJ}
\end{figure}

\begin{figure}
\centering
\includegraphics[width=\hsize]{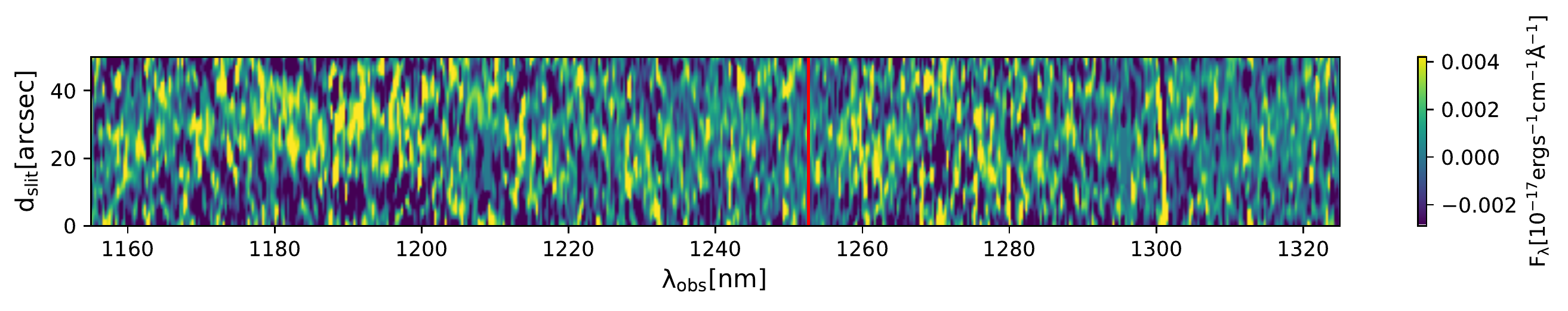}
\includegraphics[width=\hsize]{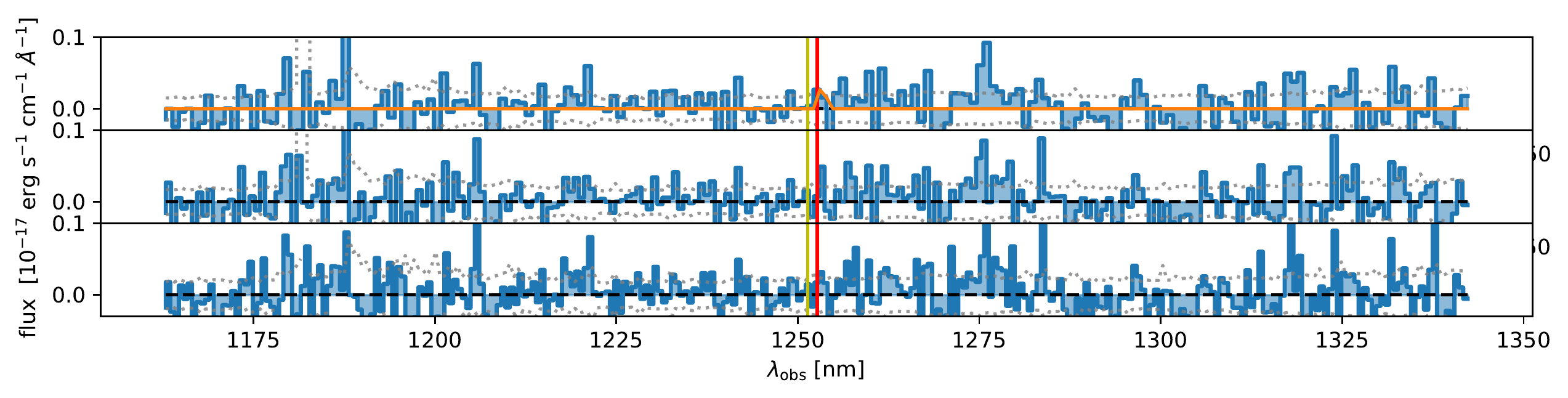}
 \caption{X-SHOOTER NIR spectra of the GDS3073 galaxy. Top panel shows the combined, cropped, and smoothed ($\rm{0.7 \ nm \times 0.12 \arcsec}$) 2D spectrum. The bottom three panels illustrate the extracted 1D spectra rebinned by $\sim 180\ \rm{km \ s^{-1}}$,  $\sim 145\ \rm{km \ s^{-1}}$, and $\sim 120\ \rm{km \ s^{-1}}$, respectively. Expected positions of the \ion{[C}{III]} $\lambda 1907$ at  $\lambda = 1251.4$ nm and \ion{C}{III]} $\lambda 1909$ line at  $\lambda = 1252.7$ nm, based on the Ly$\alpha$ spectroscopic redshift, are shown as yellow and red lines, respectively. Labels are the same as in Fig. \ref{2D-HZ1}.
          }
     \label{2D-GDS}
\end{figure}

\begin{figure}
\centering
\includegraphics[width=\hsize]{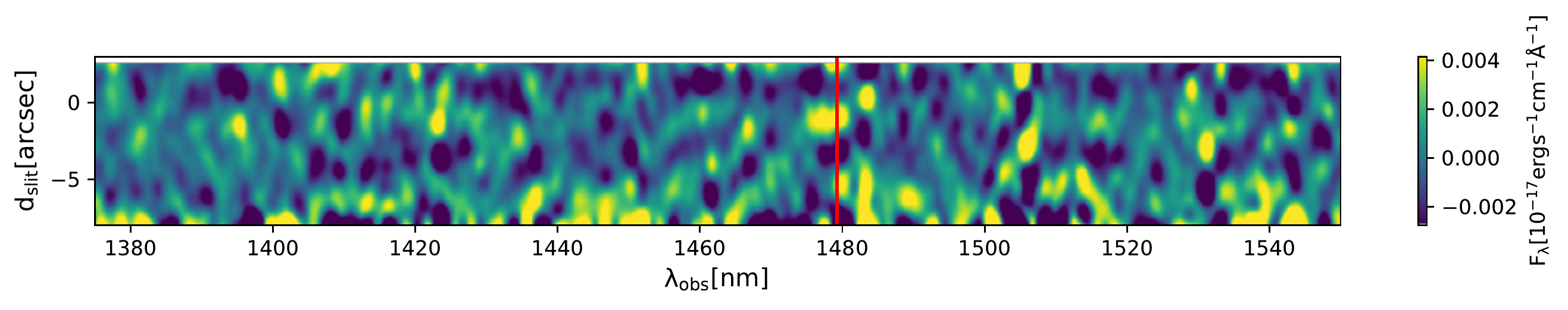}
\includegraphics[width=\hsize]{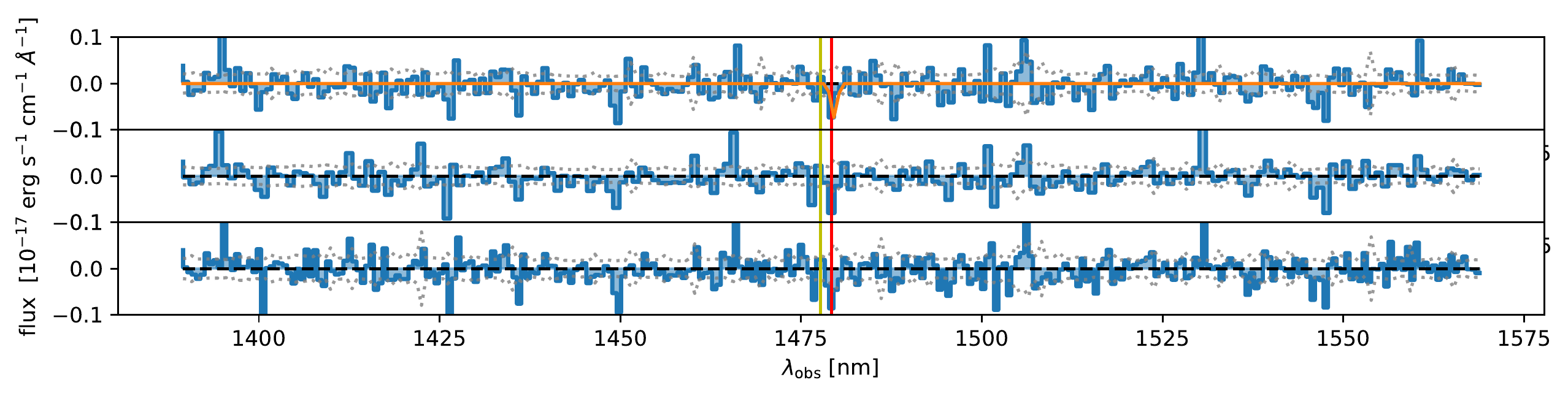}
 \caption{X-SHOOTER NIR spectra of the UVISTA-Z-002 galaxy. Top panel shows the combined, rebinned ($\sim 75 \ \rm{km \ s^{-1}}$), and smoothed ($\rm{0.7 \ nm \times 0.12 \arcsec}$)  2D spectrum. The bottom three panels illustrate the extracted 1D spectra rebinned by $\sim 180\ \rm{km \ s^{-1}}$,  $\sim 145\ \rm{km \ s^{-1}}$, and $\sim 120\ \rm{km \ s^{-1}}$, respectively. Expected positions of the \ion{[C}{III]} $\lambda 1907$ at  $\lambda = 1477.7$ nm and \ion{C}{III]} $\lambda 1909$ line at  $\lambda = 1479.2$ nm, based on the Ly$\alpha$ spectroscopic redshift, are shown as yellow and red lines, respectively.
          }
     \label{2D-UVISTA}
\end{figure}

In Figs. \ref{2D-HZ1}, \ref{2D-RXJ}, \ref{2D-GDS}, and \ref{2D-UVISTA}, we show the X-SHOOTER NIR spectra of the observed $z>5.5$ galaxies with \ion{C}{III]} line non-detection. We show the combined, rebinned, and smoothed 2D NIR spectra (top panels) and the extracted 1D spectra rebinned in wavelength by 15, 12, and 10 spectral channels, respectively (bottom panels). Shaded blue regions indicate the flux, dotted lines indicate the uncertainties, whereas orange solid line is the best Gaussian fit on the 1D spectra rebinned by 15 channels. Expected positions of the \ion{[C}{III]} $\lambda 1907$ and \ion{C}{III]} $\lambda 1909$ line, based on the Ly$\alpha$ spectroscopic redshift, are shown as yellow and red lines, respectively. 

\begin{figure}
\centering
\includegraphics[width=\hsize]{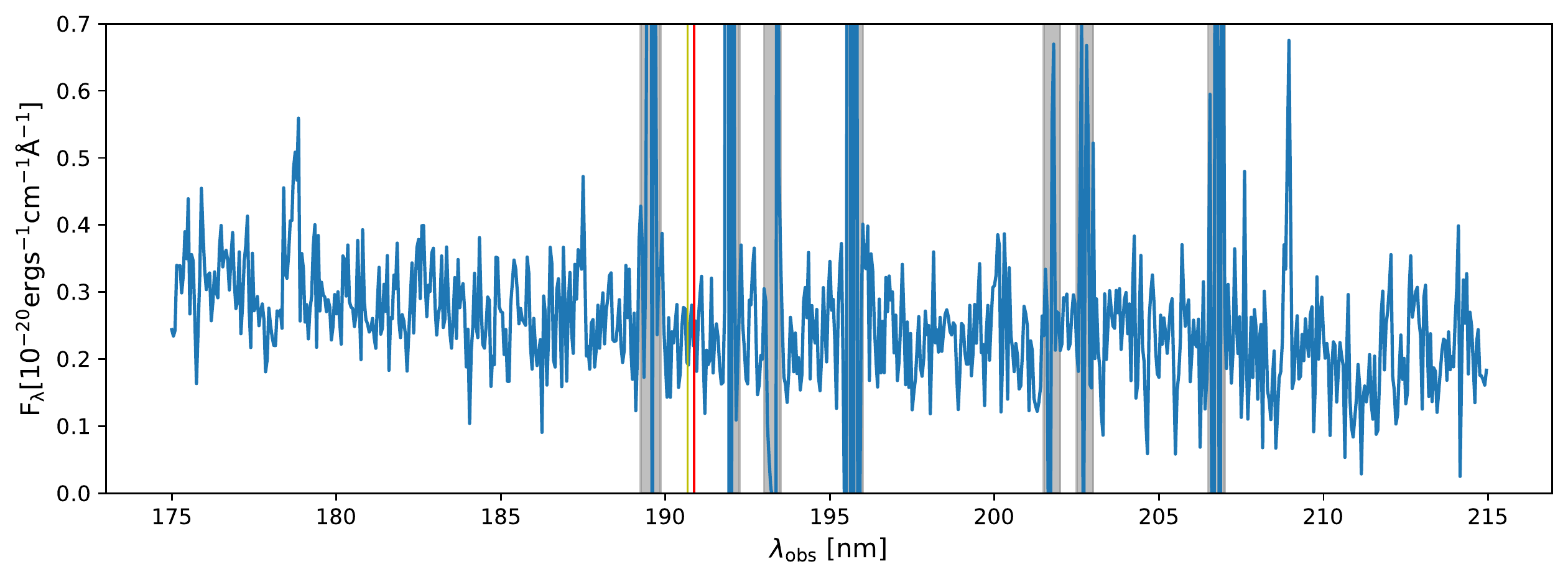}
 \caption{Stacked 1D spectrum of six $z \sim 2$ galaxies from the MUSE sample. The stacked spectrum is the weighted average of the rest-frame UV spectra of the six galaxies. The spectral resolution is set to $\Delta \lambda = 0.05$ nm. Rest-frame positions of the \ion{[C}{III]} $\lambda 1907$ and \ion{C}{III]} $\lambda 1909$ line are shown as yellow and red lines, respectively. Wavelengths with skyline contamination are shown as vertical gray regions.
          }
     \label{1D-MUSE}
\end{figure}

In Fig. \ref{1D-MUSE}, we show stacked 1D spectrum of six $z \sim 2$ galaxies from the MUSE sample. The stacked spectrum is the weighted average of the rest-frame UV spectra of the six $z \sim 2$ galaxies. The 1D spectra are rebinned to a have identical spectral resolution set to $\Delta \lambda = 0.05$ nm. We partially remove the skyline emission from the 1D spectra by subtracting the individual 1D galaxy spectrum by the 1D spectrum extracted from a $r =5\arcsec$ region close to the galaxy. Wavelengths with skyline contamination are shown as vertical gray regions.

\section{Posterior distributions of the ISM parameters}

We report the corner plot (\citealt{corner}) that illustrates the 1D and 2D projections of the posterior probability distribution of the ISM parameters: $\log{\rm{(n[\rm{cm^{-3}}])}}$, $\log{(Z[Z_{\odot}])}$, and $\log{(\kappa_s)}$, for the HZ10 galaxy. For the total gas density $\log{\rm{(n)}}$ parameter, we set an upper limit on the prior $\log{\rm{(n[\rm{cm^{-3}}])= 3.5}}$,  namely, a value close to the critical density of the emitting gas (e.g., \citealt{2019MNRAS.489....1F}). Although it might be possible for the MCMC walkers to go beyond this limit, sources with such high mean density are not very probable, because this would imply that the average density of the galaxy is much higher than the typical average density of $n \approx 300 \ \rm{cm^{-3}}$ found for high-$z$ mock galaxies from the SERRA simulations (\citealt{2019MNRAS.487.1689P}).  

 \begin{figure}
\centering
\includegraphics[width=\hsize]{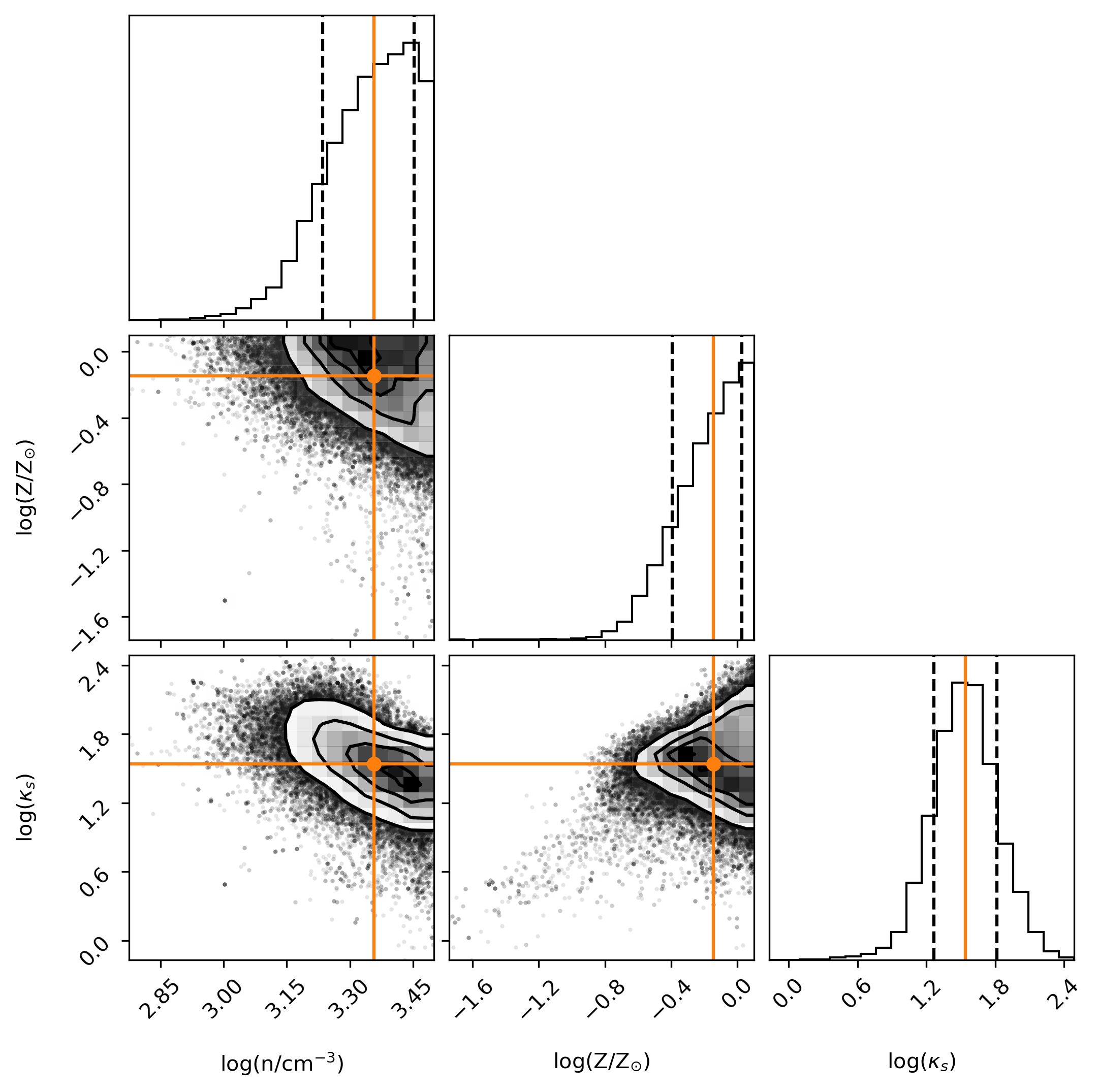}
 \caption{Corner plot representation of the posterior distribution
of the ISM parameters: $\log{\rm{(n[\rm{cm^{-3}}])}}$, $\log{(Z[Z_{\odot}])}$, and $\log{(\kappa_s)}$ for the HZ10 source.  The median of the posterior distribution and 1$\sigma$ uncertainties are represented as orange solid and black dashed lines, respectively.
          }
     \label{corner_plot}
\end{figure}

\section{ISM properties of sources with upper limits on carbon lines} \label{app_params}

Most of the sources of our sample have upper limits on either the \ion{C}{III}] or [\ion{C}{II}] emission. To be able to derive their ISM properties using the \cite{2020MNRAS.495L..22V} model, we need to treat the upper limits on the carbon line as measured values and we assume that the uncertainties are equal to $25\%$ of the upper limit estimates. With this procedure, we derive the $\kappa_s$, $Z$, and $n$ parameters. The derived values for these sources are given in Table \ref{params2}.

\begin{table}[h]
 \caption[]{\label{params2}Burstiness parameter $\kappa_s$, gas metallicity $Z$, and  density $n$ for galaxies with upper limits on either \ion{C}{III}] or [\ion{C}{II}] emission.}
 \centering
\begin{tabular}{lcccccccc}
 \hline \hline
  \noalign{\smallskip}
  ID & $\log{(\kappa_s)}$ & $Z$ & $\log{(n)}$ \\
    \noalign{\smallskip}
  & & $(Z_{\odot})$ & $(\rm{cm^{-3}})$ \\
  \noalign{\smallskip}
 \hline
 \noalign{\smallskip}
HZ1 & $0.68 $& $0.53$ & $3.07$\\
 \noalign{\smallskip}
RXJ1347-1216 & $1.34$& $0.31$ & $2.63$ \\
 \noalign{\smallskip}
GDS3073  & $2.32$&  $0.68$ & $3.45$\\ %
 \noalign{\smallskip}
 UVISTA-Z-002 & $1.40$&  $0.20$ & $2.03$\\ %
 \noalign{\smallskip}
z7\_GND\_42912  & $1.67$&  $0.58$ & $3.13$\\ %
 \noalign{\smallskip}
 Himiko  & $0.85$&  $0.19$ & $2.45$\\ %
 \noalign{\smallskip}
CR7 & $1.91$&  $0.25$ & $2.90$\\ %
 \noalign{\smallskip}
 CR7a  & $1.86$&  $0.32$ & $2.86$\\ %
 \noalign{\smallskip}
 CR7b  & $1.41$&  $0.41$ & $2.85$\\ %
 \noalign{\smallskip}
 CR7c  & $1.46$&  $0.29$ & $2.84$\\ %
 \noalign{\smallskip}
COSMOS13679  & $1.10$&  $0.27$ & $2.33$\\ %
 \noalign{\smallskip}
 COS-2987  & $0.71$&  $0.27$ & $2.35$\\ %
 \noalign{\smallskip}
 VR7  & $1.27$&  $0.38$ & $2.76$\\ %
 \noalign{\smallskip}
 9347  & $1.02$ &  $0.23$ & $2.43$\\ %
 \noalign{\smallskip}
 6515  & $1.19$&  $0.34$ & $2.70$\\ %
 \noalign{\smallskip}
 10076  & $1.42$&  $0.44$ & $2.90$\\ %
 \noalign{\smallskip}
 9834  & $2.26$&  $0.51$ & $3.40$\\ %
 \noalign{\smallskip}
 9681  & $1.46$&  $0.46$ & $2.89$\\ %
 \noalign{\smallskip}
 8490  & $1.44$&  $0.38$ & $2.80$\\ %
 \noalign{\smallskip}
\hline
\end{tabular}
\end{table}

\end{appendix}
\end{document}